\documentclass[12pt]{article}
\usepackage[T1]{fontenc}
\usepackage[english]{babel}
\usepackage[latin1]{inputenc}
\usepackage[dvips]{graphicx}
\usepackage[dvips]{epsfig}
\usepackage{amssymb}
\usepackage{color}

\def\dalemb#1#2{{\vbox{\hrule height.#2pt
        \hbox{\vrule width.#2pt height#1pt \kern#1pt \vrule width.#2pt}
        \hrule height.#2pt}}}

\def\ba{\begin{eqnarray}}
\def\ea{\end{eqnarray}}
\def\be{\begin{equation}}
\def\ee{\end{equation}}

\def\gtorder{\mathrel{\raise.3ex\hbox{$>$}\mkern-14mu
             \lower0.6ex\hbox{$\sim$}}}
\def\ltorder{\mathrel{\raise.3ex\hbox{$<$}\mkern-14mu
             \lower0.6ex\hbox{$\sim$}}}

\title{
Dissipation and nonlocality in a general expanding braneworld universe
}

\author{
Mathieu Remazeilles\thanks{E-mail: remazeil@th.u-psud.fr} \\
Laboratoire de Physique Th\'eorique,
Universit\'e Paris-Sud, 91405 Orsay, France
}

\date{}

\include{Acknowledge}

\begin{document}
\maketitle

\begin{abstract}
{ We study the evolution of both scalar and tensor cosmological perturbations
  in a Randall-Sundrum braneworld having an arbitrary expansion history. We
  adopt a four dimensional point of view where the degrees of freedom on the
  brane constitute an open quantum system coupled to an environment composed
  of the bulk gravitons. Due to the expansion of the universe, the brane
  degrees of freedom and the bulk degrees of freedom interact as they
  propagate forward in time. Brane excitations may decay through the emission
  of bulk gravitons which may escape to future infinity, leading to a sort of \emph{dissipation} from the
  four dimensional point of view of an observer on the brane. Bulk
  gravitons may also be reflected off of the curved bulk and reabsorbed by the
  brane, thereby transformed into quanta on the brane, leading to a sort of \emph{nonlocality} from the
  four dimensional point of view. The dissipation and the nonlocality are
  encoded into the retarded bulk propagator. We estimate the dissipation rates of the
  bound state as well as of the matter degrees of freedom at different
  cosmological epochs and for different sources of matter on the
  brane. We use a near-brane limit of the bulk geometry for the study when purely nonlocal bulk effects are encountered. }
\end{abstract}

\section{Introduction}

Randall and Sundrum introduced a braneworld model with an infinite
extra dimension \cite{rs} where our observable universe is a four dimensional
hypersurface, the brane, embedded into a five dimensional Anti-de Sitter
($AdS$) bulk spacetime. The curved geometry of $AdS$ space induces an
effective compactification in the extra dimension through its curvature radius
$\ell$  and thus prevents four dimensional (4d) gravity from leaking into the
extra dimension. Consequently, standard 4d Einstein gravity is recovered at
large distances on the
brane. Corrections to the Newton square inverse force law arise at scales smaller than $\ell$
\cite{gt} (in current experimental tests of Newton's law, $\ell \lesssim 0.1~
mm$ \cite{exp}). Spatially homogeneous and isotropic cosmological background
solutions in the Randall-Sundrum (RS) model have been found \cite{bdel}. These
solutions include an additional term quadratic in the
energy-momentum tensor in the Friedmann equation, causing significant
modifications of the standard cosmology at high density. In the language of
string theory, all the Standard Model fields are confined to the brane. Only
the gravitons propagate in the entire bulk spacetime. It is hoped that the
presence of an infinite extra dimension may be tested by cosmological observations, such as the measurements of temperature and polarization anisotropies of the Cosmic Microwave Background (CMB).

Cosmological perturbation theory about the cosmological background is an
essential tool to probe the presence
of extra dimensions. In braneworlds, cosmological perturbation theory is more complicated than in standard 4d cosmology
because the brane breaks the 5d spatial homogeneity and isotropy of the bulk.
Consequently, the linearized equations are no longer ordinary differential
equations (ODE) in time but become partial differential equations (PDE) in time and
the extra dimension. Analytic solutions to cosmological perturbations
have however been found for the maximally symmetric cases
of a Minkowski brane or a pure de Sitter brane \cite{rs}, \cite{lan}. In these special cases of
higher symmetry the spectrum of 5d
gravitons includes one discrete mode bound to the brane (if the bulk is $Z_2$-symmetric) and a
continuum of free modes. From the point of view of an observer on the brane, the bound
state is interpreted as the 4d massless graviton and reproduces standard 4d
Einstein gravity on the brane. The continuum is interpreted as 4d massive
Kaluza-Klein (KK) gravitons whose interaction with the brane degrees of freedom is
suppressed at low energy.  In order to explore observational
signatures for the presence of extra-dimensions, it is necessary to compute
the evolution of cosmological perturbations in the more realistic case of a
brane having a time-dependent stress-energy, \emph{i.e.} a
Friedmann-Robertson-Walker (FRW) brane, for instance with an inflation era
followed by a
radiation/matter dominated era. Several formalisms exist for evolving linearized perturbations in a
braneworld scenario \cite{pert}. Here we use the formalism of Mukohyama \cite{pert2}. The main
difficulty is that, except for the case of an uniformly accelerated brane
(flat brane or de Sitter brane)
where analytical solutions have been obtained, the equations are generally not
separable because of the complicated motion of the expanding brane in the
bulk (a review on cosmological perturbations in expanding braneworlds
is given in reference \cite{der}).
This problem has been investigated numerically and a number of interesting results have been obtained 
\cite{num0,num1,scalinf,num2,num3,num4,num5,scal,scalinf2}. The
braneworld perturbations problem is in principle entirely solvable numerically once the
initial vacuum in $AdS$ is known. However the specification of initial conditions in the bulk poses a problem of more fundamental nature. There is no unique physically motivated choice of initial
conditions for an $AdS$ bulk background \cite{trodden,bucher} (quite unlike the situation for $dS$ space). Consequently the results of these numerical studies are subject to certain assumptions concerning the initial conditions.

Five dimensional transverse and traceless metric fluctuations describe
the true bulk gravitational waves (5d gravitons) and contain five degrees of
freedom (d.o.f). By "projecting" the five d.o.f onto a flat static brane, they
split into tensor perturbations (the two d.o.f of the 4d graviton), vector
perturbations (the two d.o.f called "graviphotons") and scalar perturbations (one
d.o.f called "graviscalar"), in a $SO(3)$ representation. This "projection" is
somewhat ambiguous in the case of an arbitrary motion of the cosmological
brane. In this case it is more convenient to use directly the $SO(3)$ symmetry of the
unperturbed spacetime to decompose the linear perturbations into tensor,
vector and scalar perturbations. The advantage of this decomposition is that
scalar, vector and tensor perturbations evolve independently at linear
order. Tensor perturbations are transverse and traceless and are the easiest
to deal with. They correspond to 4d gravitational waves and propagate freely
into the bulk, independently of the presence of matter on the brane. By
definition tensor perturbations are automatically gauge-invariant at linear
order. Their wave function obeys an equation similar to a Klein-Gordon
equation for a massless scalar field minimally coupled to gravity. The
Israel junction conditions reduce merely to Neumann boundary conditions on the
brane. Vector and scalar metric perturbations are more complicated to deal with
because of their coupling to matter on the brane. Mukohyama has shown that, in the absence of matter in the
bulk, scalar (as well as vector) perturbations in the "5d longitudinal" gauge can
be derived from a single scalar master field, which obeys a five
dimensional wave equation \cite{pert2} (see
also \cite{jap}). For vector perturbations, the boundary conditions for the master field are
Dirichlet boundary conditions. In the more
interesting case of scalar perturbations, the master field satisfies more
complicated "nonlocal" boundary conditions on the brane\footnote{The term
  ``nonlocal'' for the boundary conditions has nothing to do with the
  nonlocality studied in this paper which is due to the information coming
  from the bulk. As it was stressed in \cite{deff}, the terminology ``nonlocal''
  for the boundary conditions is not appropriate, although it is frequently used
  in the literature, because they contain only a finite number
of derivatives.}. Some authors have shown
the connection between the master field formalism and the formalisms in
different gauges \cite{bridg}. Although there has been progress in
braneworld cosmological perturbation theory, there are
a few quantitative predictions for the evolution of cosmological perturbations in expanding braneworlds.

In this paper we study the evolution of cosmological perturbations
of a FRW brane having an arbitrary motion in the $AdS$
bulk. The goal of this work is to estimate the order of magnitude of
the extra dimensional effects. Instead of looking for exact solutions to the
perturbation equations, we are primarily interested in the role of the bulk
gravitons in the brane-bulk interaction, following the ideas developed in
the article of Binetruy, Bucher and Carvalho \cite{bbc}.  When the Hubble
parameter on the brane changes with time,
the acceleration of the brane in the bulk also changes and gravitons are
emitted into the bulk. These may escape to future infinity or they may be
reabsorbed by the brane because of reflections of the emitted gravitons off of
the curved $AdS$ bulk.  From
the point of view of an observer on the brane, these processes appear to generate
\emph{dissipation} and \emph{nonlocality} \cite{bbc}.  Here these bulk
induced effects are encoded into the retarded bulk propagator,
which is systematically inserted into the effective brane propagator by doing a
resummation of the bulk backreaction effects at all order in the brane-bulk
coupling. The approach used in this work is a four dimensional perspective
that regards the degrees of freedom localized on the brane universe as an open
quantum system coupled to a large environment composed of the bulk
gravitons. In the language of non-equilibrium quantum field theory, the bulk
propagator plays the same role as a self-energy, dressing
the bare fields composed of the discrete degrees of freedom localized on the brane. In this work we estimate
the magnitude of the decay rate of the graviton bound state by studying the
evolution of tensor perturbations in the frame of a cosmological brane
with an arbitrary expansion history embedded in a $Z_2$-symmetric $AdS$
space. Some authors  obtained analytical results for the order of
magnitude of the modified power spectrum for tensor perturbations, in the
simplified braneworld model where the Hubble factor on the brane changes
instantaneously \cite{pert3}. Here we consider a more
realistic model where the Hubble factor changes \emph{continuously} and
\emph{adiabatically} in the sense that $\dot{H} \ll H^2$. We also work out the dissipation of purely localized
brane degrees of freedom, such as an adiabatic perfect fluid or a slowly
rolling scalar field on the brane, by studying the evolution of scalar
cosmological perturbations and the coupling between
metric and matter perturbations. We use Gaussian normal (GN)
coordinates, where the extra dimensional coordinate measures
the proper distance from the brane and the position of the brane in
the bulk is fixed. GN coordinates have also the advantage of
simplifying the form of the perturbed boundary conditions on the brane. The main drawback
of GN coordinates is the presence of coordinate singularities in the bulk at
a finite distance from the brane, despite the bulk space being regular and
extendable beyond the singularity by
choosing another set of coordinates. Because of the arbitrary motion of a FRW
brane in $AdS$ space, the metric components in GN coordinates have a
complicated form which renders the equations of motion not separable. That is
why we use a near-brane limit to perform the separation, actually only when
purely nonlocal interactions arise between the bulk degrees of
freedom and the brane degrees of
freedom that are considered. This limit can be legitimized in the Randall-Sundrum model
because the support of the wave function of the bound state is localized near the brane. We think that physics near
the brane is suitable to describe dissipative effects in braneworlds. In order
to focus our attention on the dissipative effects, we may approximate at high
energy ($H\ell\gg 1$) the inhomogeneity of the bulk responsible for
backscattering of gravitons in the bulk and subsequent nonlocal effects on the brane. This approximation is discussed in section \ref{subsec:spectrum}.

The paper is organized as follows. In section \ref{sec:cosmo} we introduce the
dissipative and nonlocal effects which appear in braneworld cosmology and we
present the self-energy approach used in the paper to study these effects. In
section \ref{sec:bcp} we briefly recall the cosmological perturbation theory in
braneworlds. In section \ref{sec:scalar2} we explore the scalar
perturbations and the brane-bulk interaction for a slowly rolling scalar
field on the brane. We apply our method to compute the local dissipation rate
of the inflaton perturbation due to the interaction with the bulk. In section
\ref{sec:metric} we compute the Gaussian normal metric of $AdS$ space with
general Friedmann-Robertson-Walker (FRW) slices and present the near-brane
limit of this geometry, which will be used as an approximation to evaluate the
nonlocal corrections arising in the brane-bulk systems considered in the next
sections. In section \ref{sec:eom} we study the tensor perturbations and
compute the dissipation rate of the graviton bound state. In section
\ref{sec:scalar} we study the scalar perturbations for an adiabatic perfect
fluid on the brane. We present our conclusions in section \ref{sec:conclusion}.

\section{Dissipation and nonlocality in the expanding braneworld}\label{sec:cosmo}

\begin{figure}[htbp]
\setlength{\unitlength}{1cm}
\begin{center}
\begin{minipage}[t]{5.cm}
\begin{picture}(5.5,5.5)
\centerline {\hbox{\psfig{file=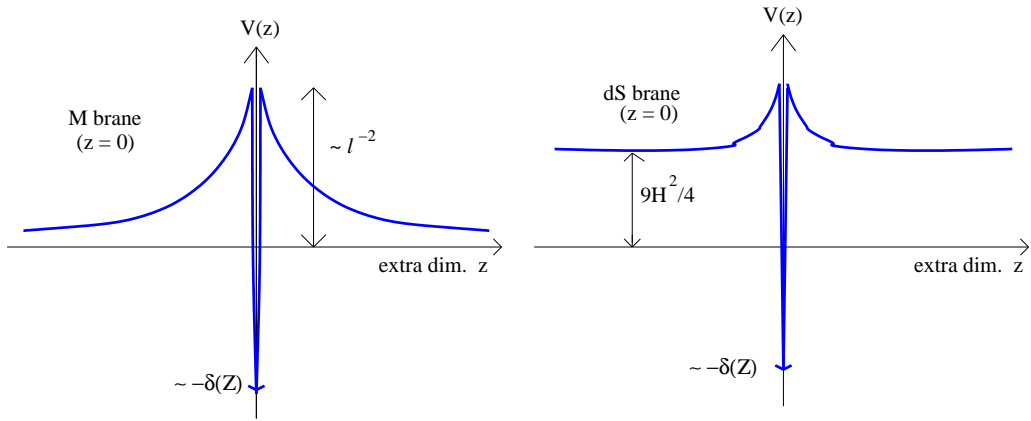,height=5.5cm}}}
\end{picture}\par
\end{minipage}
\hfill
\end{center}
\caption{\small{{\bf Left: Volcano potential for a Minkowski braneworld.} The height of
  the potential is $\mathcal{O}(\ell^{-2})$ and the potential decreases
  as $1/z^2$ if $z$ is the extra dimension. The zero mode graviton bound to
  the brane is surrounded by a continuum of
  massive gravitons $m > 0$. {\bf Right: Volcano potential for
  a de Sitter braneworld.} The height of
  the potential is $\mathcal{O}(H^2+\ell^{-2})$ and the potential decreases
  as $1/\sinh^2(z)$.The zero mode graviton bound to the brane is surrounded by a continuum of
  massive gravitons with $m > 3H/2$ characterized by a gap.}} \label{fig:volcano}
\end{figure}

Cosmological perturbations theory for the Randall-Sundrum model has been
solved analytically for special brane universes where the Hubble factor on
the brane $H$  is constant. In these cases the four dimensional universe is
static ($H = 0$)  or follows a uniform expansion ($H > 0$). In the first case
the Minkowski brane is at rest with respect to the five dimensional Anti-de
Sitter ($AdS$) bulk, whereas in the second case the pure de Sitter ($dS$)
brane follows a uniformly accelerating trajectory  in the $AdS$ bulk. The
equations for the perturbations may be reduced to a Schr\"odinger-like equation. The presence of the brane within the
bulk creates a $\delta$-function potential well, while the curvature of the
$AdS$ bulk creates a decreasing barrier potential (Fig. \ref{fig:volcano}).
The resulting volcano-like potential gives rise to a single brane
bound mode, the zero mass mode $m = 0$,  in the spectrum of the linearized
Einstein equations, surrounded by a continuum of massive Kaluza-Klein free
modes, with $m \geq 3H/2$ for a $dS$ brane \cite{lan}. The zero mode
represents the standard massless graviton and reproduces the four dimensional
gravity on the brane. The amplitude of the massive gravitons is suppressed near the brane at low energy due to the barrier potential.

Because of the presence of matter in the universe, the cosmological expansion
history is no longer uniform. The Hubble parameter changes with time, inducing
a change in the shape of the potential.
Consequently, the brane degrees of freedom and the bulk degrees of freedom
interact thus generating transitions between the modes.

The brane-bulk system is a Hamiltonian quantum system, which is necessarily
conservative because the phase space density must be preserved under time
evolution. In this sense, the brane-bulk system is intrinsically non-dissipative, and the appearance of dissipation can arise only as the result of
``coarse graining''. The interaction between the quanta due to the expansion
of the brane may be characterized by a Bogoliubov transformation ($S$ matrix
for a linear system) relating the modes of \emph{positive} and \emph{negative}
frequency for the initial ``in'' vacuum and the resulting ``out'' vacuum. However,
from the point of view of an observer confined to the brane, the four
dimensional universe as a submanifold  is an open quantum
system. Consequently, the mixing between the modes localized on the brane and
the delocalized modes in the bulk creates the appearance of dissipation to a
four dimensional observer unable to access the bulk modes. The observer on the
brane is sensitive only to a part of the complete $AdS$ vacuum,
composed of only the discrete modes localized on the brane. We call this part
``brane vacuum''. The Hilbert space is truncated
because of dimensional reduction, and the ``bulk vacuum'', composed
of the continuous bulk modes, holds the missing information. When we observe
the cosmological perturbations today, we measure expectation values of
observables quadratic in the creation and annihilation operators for the modes
localized on the brane today, namely $a_{brane,out}$ and
$a^\dagger_{brane,out}$. The $S$ matrix expresses the ``out'' operators as
linear combinations of $a_{brane,in}$ and $a^\dagger_{brane,in}$ on one hand,
and of $a_{bulk,in}(k)$ and $a^\dagger_{bulk,in}(k)$ on the other. A useful parameterization of this transformation
has been proposed in \cite{bbc}. The authors require that $A_{brane,in}$ and
$A_{bulk,in}$ be entirely on the brane and in the bulk, respectively, and be normalized
such that $\left[A_{brane,in},A^\dagger_{brane,in}\right] =
\left[A_{bulk,in},A^\dagger_{bulk,in}\right] = 1$. Then $a_{brane,out}$ may be expressed in terms of these according to one of the three following possibilities: either
\ba
a_{brane,out} & = & \cos\theta A_{brane,in}+\sin \theta A_{bulk,in}
\ea
where $0\leq\theta\leq\pi/2$; or
\ba
a_{brane,out} & = & \cosh u A_{brane,in}+\sinh u A^\dagger_{bulk,in}
\ea
where $0\leq u\leq +\infty$; or
\ba\label{qqq:icp}
a_{brane,out} & = & \sinh u A^\dagger_{brane,in}+\cosh u A_{bulk,in}
\ea
where $0\leq u\leq +\infty$. $A_{brane,in}$ may be constructed entirely as a
linear combination of $a_{brane,in}$ and $a^\dagger_{brane,in}$, and likewise
$A_{bulk,in}$ may be constructed entirely as a linear combination of the
$a_{bulk,in}(k)$ and the $a^\dagger_{bulk,in}(k)$, $k$ being the index for the
continuum modes.
 We observe in (\ref{qqq:icp}) that the bulk initial state
may have a very important, or even dominant, role in determining what we see
on the brane today.

However it has been numerically observed in \cite{num4} that the evolution of
tensor perturbations on the brane today does not show any significant
dependence on the bulk initial state when a $dS$-invariant initial vacuum
is specified in the $AdS$ bulk, that is defined in a $dS$-slicing frame. However we might suspect that the initial amplitude
of bulk modes that are initially located outside the bulk Cauchy horizon of the
$dS$-slicing will not be suppressed when they causally affect the decelerating brane in the
future. The relative contribution of the
brane initial conditions and the bulk initial conditions in determinining what we see on the
brane today should depend on the basis decomposition of the \emph{in}-vacuum as described
in (\ref{qqq:icp}). The main problem is that there are no natural initial conditions
for $AdS$ \cite{trodden,bucher}: in standard inflation the backgound geometry
is $dS$ so the timelike geodesics diverge forward in
time losing causallity such that initial
irregularities become swept out, $dS$ space thus exhibits natural initial
conditions for generating homogeneity and isotropy. Whereas in Randall-Sundrum
branewords the background geometry is $AdS$ so that the timelike
geodesics first diverge and then refocus forward in time, thus conserving the
initial amplitude of metric perturbations.

\begin{figure}[htbp]
\setlength{\unitlength}{1cm}
\begin{center}
\begin{minipage}[t]{5.cm}
\begin{picture}(5.5,5.5)
\centerline {\hbox{\psfig{file=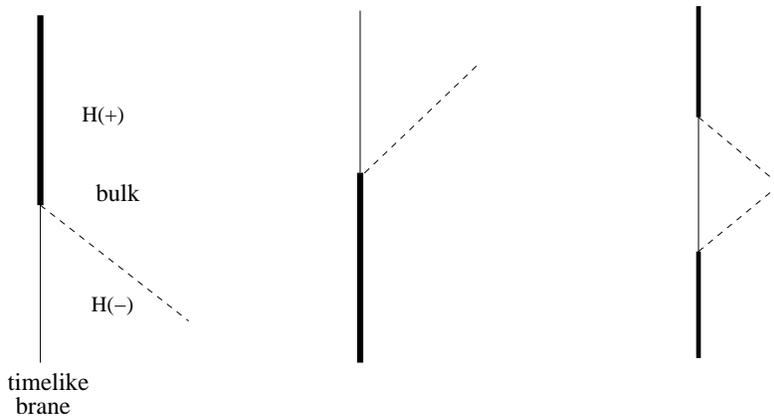,height=5.5cm}}}
\end{picture}\par
\end{minipage}
\hfill
\end{center}
\caption{\small{{\bf Fundamental processes for the brane-bulk interaction from a four
  dimensional perspective.} From the left to the right: absorption,
  dissipation, nonlocality. The nonlocality arises from the
  diffraction of gravitons into the curved $AdS$ bulk.}}\label{fig:interaction}
\end{figure}

The brane-bulk interaction may be summarized by the following fundamental
processes from the four dimensional point of view illustrated in  Fig. \ref{fig:interaction}. An initial vacuum state may
be completely characterized by specifying the quantum state of the incoming
gravitons on the past Cauchy horizon in the bulk $H_{(-)}$ and of the degrees
of freedom on the brane at the intersection of the
brane with $H_{(-)}$. Subsequently, the bulk and the brane degrees of freedom
interact as they propagate forward in time.  Bulk gravitons may be absorbed
and transformed into quanta on the brane. Similarly, brane excitations may
decay through the emission of bulk gravitons. These may either escape to
future infinity, leading to \emph{dissipation} from the four dimensional point
of view,
or may be reabsorbed by the brane due to the bulk curvature,
leading to
\emph{nonlocality} from the four dimensional point
of view.

We may consider the brane-bulk interaction from the four dimensional perspective by regarding the brane degrees of freedom as an open quantum system
coupled to an environment with a large number of degrees of
freedom. Schematically, we may block diagonalize by Fourier transforming in the
three transverse spatial dimensions. For an expanding braneworld with a
perfect fluid or scalar field on the brane, the
interaction between the scalar brane degree of freedom $q(t)$ and a scalar bulk degree of
freedom $u(t,x)$ may be represented by a system of coupled equations. Under
the approximations presented in detail later in the paper, this system of coupled equation may be written according to the general form
\ba
\left[-\partial_t a(t,x) \partial_t + \partial_x b(t,x) \partial_x + c(t,x)\right] u(t,x) & = & 0,\cr
\left[\partial_x +\lambda(t)\right] u(t,x)\vert_{x = 0} & = & P_1(\partial_t)q(t),\cr
\left[\partial_t^2+\omega^2_0(t)\right]q(t) & = & P_2(\partial_t)u(t,x)\vert_{x = 0},
\ea
where $P_1$ and $P_2$ are polynomials in $\partial_t$ of degree one and degree
two respectively, with time dependent coefficients. The first equation is the
bulk equation of motion, the second the
boundary condition on the brane, and the third the equation of motion for the
brane degree of freedom. The brane and the bulk degrees of freedom are related
according to
\ba
q(t) & = & \int dt' G_{brane}(t,t') P_2(\partial_{t'}) u(t',0),\cr
u (t,x) & = & \int dt' G_{bulk}(t,t',x,0) P_1(\partial_{t'}) q(t'),
\ea
where $G_{brane} = \left[\partial_t^2+\omega^2_0(t)\right]^{-1}$ is the
bare\footnote{By ``bare'' we mean the propagator of the free brane fields excluding
  the interaction with the bulk.}
retarded brane propagator and $G_{bulk}$ is the Neumann form of the retarded bulk propagator. Due
to the interaction, the effective retarded brane propagator may be obtained
 by resumming the infinite geometric series at all orders of the bulk
 backreaction
\ba
\hat{G}_{brane} & = & G_{brane}+G_{brane}P_2(D_t)G_{bulk}P_1(D_t)G_{brane}\cr
                &  & +G_{brane}P_2(D_t)G_{bulk}P_1(D_t)G_{brane}P_2(D_t)G_{bulk}P_1(D_t)G_{brane}+ ...\cr
                & = & {1\over G_{brane}^{-1}-P_2(D_t)G_{bulk}P_1(D_t)},
\ea
where the bulk propagator links two points both lying on the brane. The brane
degree of freedom is governed by the integro-differential equation
\ba
\hat{G}^{-1}_{brane}\circ q & = & \partial^2_t q+\omega_0^2(t)q+\int dt'
K_{bulk}(t-t')q(t') = 0.
\ea
Here the bulk kernel is given by
\ba
K_{bulk}(t,t') = -P_2(\partial_t)G_{bulk}(t,t')P_1(\partial_{t'})
\ea
and plays the role of a self-energy, dressing the bare field on the brane. Both
the local dissipative effects and the nonlocal effects due to the interaction
with the extra
dimension are contained in this self-energy kernel. The bare retarded brane
propagator $G_{brane} = \left[\partial_t^2+\omega^2_0(t)\right]^{-1}$ is the
propagator of an  oscillator localized on the brane and accounts only for the transitions between the four dimensional modes due to the expansion of the universe.

We consider as an illustration the following couplings
\ba
P_1(\partial_t) & = & -\gamma_{1}(t) \partial_t, \qquad P_2(\partial_t)  =  \gamma_{2}(t) \partial_t.
\ea
Since the Neumann retarded bulk propagator on the brane can be written in the
form
\ba
G_{bulk}(t,t',x = 0, x' = 0) = \theta(t-t')G(t,t',0,0),
\ea
the bulk kernel may be decomposed into singular and regular parts
\ba
K_{bulk}(t,t') & = & K^{sing}_{bulk}(t,t')+K^{reg}_{bulk}(t,t')
\ea
where
\ba
K^{sing}_{bulk}(t,t') & = & \delta(t-t')\gamma_1(t)\gamma_2(t)G(t,t,0,0)\partial_t\cr
K^{reg}_{bulk}(t,t') & = & \theta(t-t')\gamma_2(t)\partial_t G(t,t',0,0)\gamma_1(t')\partial_{t'}.
\ea
The singular part is responsible for the local dissipative processes and the
regular part describes subsequent nonlocal processes, such as reflections from
the inhomogeneous curved bulk. The infinite summation
\ba
\tilde{G}_{brane} & = &
G_{brane}+G_{brane}\left(-K^{sing}_{bulk}\right)G_{brane}+G_{brane}\left(-K^{sing}_{bulk}\right)G_{brane}\left(-K^{sing}_{bulk}\right)G_{brane}+ ...,\cr
                & = & {1\over \partial_t^2+\Gamma(t)\partial_t+\omega^2_0(t)}
\ea
is equivalent to adding a local dissipation term $\Gamma(t) = \gamma_1(t)\gamma_2(t)G(t,t,0,0)$ to the equation of motion for the brane degree of freedom. The infinite summation
\ba
\hat{G}_{brane} & = &
\tilde{G}_{brane}+\tilde{G}_{brane}\left(-K^{reg}_{bulk}\right)\tilde{G}_{brane}\cr
                 & & +\tilde{G}_{brane}\left(-K^{reg}_{bulk}\right)\tilde{G}_{brane}\left(-K^{reg}_{bulk}\right)\tilde{G}_{brane}+ ...
\ea
adds the nonlocality to the equation of motion such that the brane degree of freedom propagates as
\ba
\hat{G}^{-1}_{brane}\circ q & = & \ddot{q}(t)+\Gamma(t)\dot{q}(t)+\omega^2_0(t)q(t)+\gamma_2(t)\int_{-\infty}^t dt' G(t,t',0,0)\gamma_1(t')\dot{q}(t')  =  0.
\ea
Since a local singular part does not depend on the curvature of the background, the
term $G(t,t,0,0)$ which appears in the local dissipation rate may be replaced
by the Neumann form of the \emph{Minkowski} propagator. We recall that the
Minkowski propagator is
\ba
G_{Mink}(t,t',x,x') & = &
\theta\left((t-t')^2-(x-x')^2\right)J_0\left(p\sqrt{(t-t')^2-(x-x')^2}\right),
\ea
where $p$ is the momentum in the three transverse spatial dimensions. Then
$G(t,t,0,0) = G_{Mink}(t,t,0,0) = 1$ and the local dissipation rate depends only on the
couplings
\ba
\Gamma(t) & = &\gamma_1(t)\gamma_2(t).
\ea

For certain derivative couplings, additional local terms may appear leading for
instance to a phase shift. There might also be a nonlocal dissipation term when
there is no derivative in the couplings.
 This formalism with Green functions is useful to discriminate
 transitions among four dimensional brane modes due to the expansion of
 the universe from transitions between the brane and the bulk modes due to
 the presence of an extra dimension. Moreover this formalism allows us to
 distinguish the local dissipation processes from the nonlocal processes.


The purpose of this work is to estimate the dissipation rates of
 certain degrees of freedom confined to the brane or localized near the brane
 from an arbitrary expansion history of a brane having no special symmetries.

\section{Braneworld cosmological perturbations}\label{sec:bcp}

In this section we recall briefly the cosmological perturbations equations in
braneworlds and the formalism of Mukohyama frequently used in the literature
\cite{deff,pert2,jap,bridg,scalbc,scal,scalinf,scalinf2}.
The metric perturbations $h_{\mu\nu}$ about the general bulk background metric
$g_{\mu\nu}$, in Gaussian normal coordinates, given by
\ba\label{eqnarray:typicmetric}
ds^2 = g_{\mu\nu}dx^\mu dx^\nu & = &
d\xi^2-N^2(\xi,\tau)d\tau^2+A^2(\xi,\tau)d\mathbf{x}_{3}^2,
\ea
describe the five dimensional bulk gravitons. The evolution of these
bulk perturbations (where we assume $h_{\mu\nu} \ll g_{\mu\nu}$) will depend on the coupling to the matter
perturbations $\delta T_{\mu\nu}$ localized on the brane at $\xi = 0$
through the linearized Einstein equations
\ba
\delta G_{\mu\nu}\left(h_{\mu\nu}\right) = \kappa^2 \delta
T_{\mu\nu}
\ea
from which linearized Israel junction conditions on the brane can be
derived. The coupling is related to the five dimensional Planck mass
$\kappa^2 = 8\pi/M_{pl}^3 = 8\pi G_N\ell$, where $G_N$ is the four dimensional
gravitational coupling constant and $\ell$ the curvature radius of the bulk
space in the Randall-Sundrum model.
The bulk metric perturbations can be decomposed into scalar, vector, and tensor
perturbations.

\subsection{Tensor perturbations}

The gravitational waves away from the brane propagate freely in the
bulk and independently of the presence of matter on the brane. They are described
by transverse and traceless tensor perturbations $E_{ij}$ of the metric
\ba
d\tilde{s}^2 =  \left(g_{\mu\nu}+h_{\mu\nu}\right)dx^\mu dx^\nu & = &
d\xi^2-N^2(\xi,\tau)d\tau^2+A^2(\xi,\tau)\left(\delta_{ij}+E_{ij}\right)dx^idx^j.
\ea
 Tensor perturbations are gauge invariant at linear order and they do not
 couple to matter on the brane. The tensor perturbations are exactly
described by a minimally coupled massless scalar field $\Psi(t,\xi; p)$,
Fourier transformed in the three transverse spatial dimensions, as
\ba
E_{ij}(t,\xi,{\bf x}) & = & \int dp\Psi(t,\xi; p)e^{i{\bf p}\cdot{\bf x}}e_i\otimes e_j,
\ea
 which propagates in the bulk background according to the Klein-Gordon equation
\ba\label{eqnarray:eomt} -{1\over N^2}\left[\ddot{\Psi}+\left(3{\dot{A}\over A}-{\dot{N}\over
      N}\right)\dot{\Psi}\right]+\left[\Psi''+\left(3{A'\over A}+{N'\over
      N}\right)\Psi'\right]-{p^2\over A^2}\Psi & = & 0.\ea
The Israel junction conditions reduce to a Neumann boundary
condition on the brane
\ba\label{eqnarray:bct}
\partial_\xi \Psi\bigr\vert_{\xi = 0} & = & 0.
\ea
This boundary condition is homogeneous under the assumption that there are no anisotropic stresses on the brane.

\subsection{Scalar  perturbations}\label{subsec:scalarpert}

The scalar perturbations of the bulk background metric may be simplified by
choosing a particular gauge
knowned as ``5d longitudinal gauge''. In this gauge there are four gauge-invariant variables and the line element is
\ba
d\tilde{s}^2 & = &
\left(1+2\mathcal{A}_{\xi\xi}\right)d\xi^2-N^2(\xi,\tau)\left(1+2\mathcal{A}\right)d\tau^2+N(\xi,\tau)\mathcal{A}_\xi
d\xi dt\cr
             &   &+A^2(\xi,\tau)\left(1+2\mathcal{R}\right)\delta_{ij}dx^idx^j.
\ea
These five dimensional scalar metric perturbations couple to the four
dimensional scalar matter perturbations on
the brane, which are defined in the same gauge on the brane by\footnote{We have
  neglected the scalar anisotropic stress which do not arise when considering
  the perturbations of a perfect fluid or a scalar field.}
$$
T_{\mu\nu}+\delta T_{\mu\nu}  =
\left[
\begin{array}{cc}
-(\rho+\delta\rho) & \delta q_{,j} \\
-A^{-2}\delta q_{,i} & \left(P+\delta P\right)\delta_{ij}
\end{array}
\right].
$$
Here the greek indices label the four dimensions on the brane and the latin
indices label the three transverse spatial dimensions on the brane. Throughout
the paper, $\rho$ and $P$ denote respectively the
energy density and the pressure of the \emph{effective} matter on the brane as
opposed to the \emph{real} matter on the brane labeled by a subscript $M$ and
related to the previous one by $\rho = \rho_M+\sigma$, $P = P_M-\sigma$, where
$\sigma$ is the brane tension, fine-tuned in the Randall-Sundrum model so that $\sigma =
6/(\kappa^2\ell)$ \cite{rs}.
Mukohyama \cite{pert2}
(see also \cite{jap}) has shown that in the absence of bulk
matter perturbations, the scalar
metric perturbations in 5d longitudinal
gauge can all be generated from a single scalar ``master''
field\footnote{It should be noted that $[\Omega] = (length)^2$.} $\Omega(t,\xi)$ according to
\ba
\mathcal{A} & = & -{1\over 6A}\left\{\left(2\Omega''-{N'\over N}\Omega'\right)+{1\over
    N^2}\left(\ddot{\Omega}-{\dot{N}\over
      N}\dot{\Omega}\right)-{1\over\ell^2}\Omega\right\},\cr
\mathcal{A}_\xi & = & {1\over N A}\left(\dot{\Omega}'-{N'\over N}\dot{\Omega}\right),\cr
\mathcal{A}_{\xi\xi} & = & {1\over 6A}\left\{\left(\Omega''-2{N'\over N}\Omega'\right)+{2\over
    N^2}\left(\ddot{\Omega}-{\dot{N}\over
      N}\dot{\Omega}\right)+{1\over\ell^2}\Omega\right\},\cr
\mathcal{R} & = & {1\over 6A}\left\{\left(\Omega''+{N'\over N}\Omega'\right)-{1\over
    N^2}\left(\ddot{\Omega}-{\dot{N}\over
      N}\dot{\Omega}\right)-{2\over\ell^2}\Omega\right\}.\label{eqnarray:R}
\ea
The master variable $\Omega$ satisfies the five
dimensional wave equation
\ba\label{eqnarray:eoms} -{1\over N^2}\left[\ddot{\Omega}-\left(3{\dot{A}\over A}+{\dot{N}\over
      N}\right)\dot{\Omega}\right]+\left[\Omega''-\left(3{A'\over A}-{N'\over
      N}\right)\Omega'\right]-\left({p^2\over A^2}-{1\over \ell^2}\right)\Omega & = & 0.\ea
The Israel junction conditions reduce to the following "nonlocal"
boundary conditions on the brane \cite{scalbc}
\ba\label{eqnarray:sbc0}
\kappa^2 A\delta\rho & = & -3{\dot{A}\over A}\left(\dot{\Omega}'-{N'\over N}\dot{\Omega}\right)-{p^2\over
A^2}\left(\Omega'-{A'\over A}\Omega\right)\Bigr\vert_{\xi = 0},\cr
\kappa^2 A\delta
q & = & -\left(\dot{\Omega}'-{N'\over N}\dot{\Omega}\right)\Bigr\vert_{\xi = 0},\cr
\kappa^2 A\delta
P & = &
\left(\ddot{\Omega}'-{A'\over A}\ddot{\Omega}\right)+2{\dot{A}\over
  A}\left(\dot{\Omega}'-{N'\over
    N}\dot{\Omega}\right)-\left(\left(2{\dot{A}\over A}\right)'+\left({\dot{N}\over N}\right)'\right)\dot{\Omega}\cr
                 &   & +\left({A'\over A}-{N'\over N}\right)\left({1\over\ell^2}-{2\over 3}{p^2\over A^2}\right)\Omega-\left({A'\over A}-{N'\over N}\right)\left(2{A'\over A}-{N'\over N}\right)\Omega'\Bigr\vert_{\xi = 0}.
\ea

\section{Scalar perturbations: slow-roll inflaton on
the brane}\label{sec:scalar2}

In this section we work out scalar metric and matter perturbations for a
scalar field on the brane with a slow-roll potential, $V(\phi)$, such that the
induced geometry on the brane is quasi-de Sitter. During slow-roll inflation
the expansion of the universe is adiabatic in the sense that $\dot{H} \ll
H^2$. The inflaton scalar field, $\phi(\tau)$, may be characterized by an
energy density $\rho_M$ and a pressure $P_M$
\ba
\rho_M  & = & {1\over 2}\dot{\phi}^2+V(\phi),\cr
P_M & = & {1\over 2}\dot{\phi}^2-V(\phi).
\ea
We may combine the equations of Mukohyama presented in section
\ref{subsec:scalarpert} for scalar perturbations to obtain simple coupled
equations of the brane-bulk system. Following the calculations done in \cite{scalinf,scalinf2},
we introduce the Mukhanov-Sasaki variable as the brane degree of freedom:
\ba
Q  & = & \delta\phi-{\dot{\phi}\over H}\mathcal{R}_b,
\ea
where $\delta\phi$ is the perturbation of the inflaton and $\mathcal{R}_b$ is
one of the scalar metric perturbations, namely the curvature perturbation (\ref{eqnarray:R}),
projected on the brane. The equations of the brane-bulk system in the general Gaussian normal coordinate system (\ref{eqnarray:typicmetric})  are then given by \cite{scalinf,scalinf2}
\ba\label{qqq:braneinf}
 -{1\over N^2}\left[\ddot{\Omega}-\left(3{\dot{A}\over A}+{\dot{N}\over
      N}\right)\dot{\Omega}\right]+\left[\Omega''-\left(3{A'\over A}-{N'\over
      N}\right)\Omega'\right]-\left({p^2\over A^2}-{1\over \ell^2}\right)\Omega & = & 0,\cr
-{p^2\over
  a^2}\left[H\left(\Omega'-{A'\over A}\Omega\right)+{\kappa^2\dot{\phi}^2\over 6}\left(\dot{\Omega}-H\Omega\right)\right]\Biggr\vert_{\xi = 0} & = &
\kappa^2 a \dot{\phi}^2\left({H\over\dot{\phi}}Q\right)^{.},\cr
\ddot{Q}+3H\dot{Q}+ \left({p^2\over
a^2}+V''(\phi)\right)Q +\left\{{\ddot{H}\over H}-2{\dot{H}\over H}{V'(\phi)\over\dot{\phi}}-2\left({\dot{H}\over H}\right)^2\right\}Q& = & J(\Omega)\vert_{\xi = 0},
\ea
where $p$ is the transverse momentum and the source $J(\Omega)$ is given by

\ba\label{qqq:braneinf2}
J(\Omega) & = & -{\dot{\phi}\over H}\Biggl[\left({-\dot{H}\over H}+{\ddot{H}\over 2\dot{H}}\right){p^2\over 3 a^3}\left(\dot{\Omega}-H\Omega\right)+\left(1-{\dot{H}\over 2\mathcal{H}^2}\right){p^4\Omega\over 9a^5}\cr
          &   & +{p^2\over 6a^3}\left(\ddot{\Omega}-H\dot{\Omega}+{p^2\over 3a^2}\Omega-\left({N'\over N}-{A'\over A}\right)\Omega'\right)\cr
          &   & +{\dot{H}\over \mathcal{H}^2}{p^2\over 6a^3}\left(\mathcal{H}\Omega'-H\dot{\Omega}-{1\over\ell^2}\Omega+{p^2\over 3a^2}\Omega\right)\Biggr].
\ea
Here the notations are $a(\tau) = A(\tau,\xi = 0)$ and $\mathcal{H} =
(A'/A)\vert_{\xi = 0}$. From the Friedmann equation (\ref{qqq:bwdyn}) we also have
\ba
\kappa^2\dot{\phi}^2/2 & = & -\ell\dot{H}/\sqrt{1+\ell^2H^2}.
\ea

The brane-bulk equations (\ref{qqq:braneinf}),
(\ref{qqq:braneinf2}) may be simplified for slow-roll
inflation as follows: we neglect all the adiabatic corrections to the
expansion, like $\dot{H}/H^2$ terms, except
for the terms involved in the coupling between $Q$ and $\Omega$. According to
these approximations and after rescaling the master field according to
$A\Omega \rightarrow \Omega$, we argue that the system of coupled equations reduces to
\ba\label{eqnarray:lastsyst}
G_{bulk}^{-1}(\tau,\xi,\tau',\xi')\circ\Omega(\tau',\xi')  & = & 0,\cr
\Omega'\vert_{\xi = 0} & = &P_1(\partial_\tau)Q(\tau),\cr
G_{brane}^{-1}(\tau,\tau')\circ Q(\tau') & = & P_2(\partial_\tau)\Omega(\tau,\xi)\vert_{\xi = 0},
\ea
where $G_{bulk}$ is the retarded Green function of exact bulk equation of
motion, \\
\noindent
 $G_{brane}~=~\left[\partial_\tau^2+3H(\tau)\partial_\tau+p^2/a^2\right]^{-1}$
 is the bare Green function of the inflaton perturbation on the brane, $P_1(\partial_\tau)~=~-(\kappa^2 a^2\dot{\phi}/p^2)\partial_\tau$ and $P_2(\partial_\tau) =
-(p^2\dot{\phi}/(6a^2 H))\left[\partial_\tau^2+H(\tau)\partial_\tau+p^2/a^2\right]$.
The first equation in (\ref{eqnarray:lastsyst}) is the bulk wave equation.
The presence of time derivatives of the fields in the brane-bulk couplings induces
\emph{local} dissipation of the scalar field on the brane.
The effective brane propagator with the interaction with the bulk taken into
account is obtained as explained in section \ref{sec:cosmo} by resumming the
infinite geometric series
\ba\label{qqq:effpropinf}
& & \hat{G}_{brane}\cr
& = & G_{brane} + G_{brane}\left[{p^2\dot{\phi}\over
    6a^2H}\left( \ddot{G}^N_{bulk}+H\dot{G}^N_{bulk}+{p^2\over
        a^2}G^N_{bulk}\right){\kappa^2 a^2 \dot{\phi}\over p^2}D_\tau\right] G_{brane} + ...\cr
                & = & {1\over G^{-1}_{brane}-{p^2\dot{\phi}\over
    6a^2H}\left(\ddot{G}^N_{bulk}+H\dot{G}^N_{bulk}+{p^2\over
        a^2}G^N_{bulk}\right){\kappa^2 a^2 \dot{\phi}\over p^2}D_\tau}
\ea
where $G^{-1}_{brane} = \left[D_\tau^2+3H D_\tau+p^2/a^2\right]$ describes the
bare propagation on the brane. $G^N_{bulk}$ is the Neumann form of the bulk
retarded propagator projected on the brane and has the form
\ba\label{qqq:gbulkform}
G^N_{bulk}(\tau,\tau',\xi = 0,\xi' = 0) & = & \theta(\tau-\tau')G^N(\tau,\tau',0,0).
\ea
The time derivatives of the bulk propagator (\ref{qqq:gbulkform}) appearing in the effective brane
propagator (\ref{qqq:effpropinf}) lead to singular and regular terms.
This suggests that the dressed propagation of the inflaton contains local and
nonlocal adiabatic corrections

\ba\label{qqq:geqscal}
 &  & \left[1-{\kappa^2\dot{\phi}^2\over 6H}G^N(\tau,\tau,0,0)\right]\ddot{Q}+\left[3H-{\kappa^2\dot{\phi}^2\over 6H}\left(HG^N(\tau,\tau,0,0)+\dot{G}^N(\tau,\tau,0,0)\right)\right]\dot{Q}\cr
& & +{p^2\over a^2}Q+\mbox{ (nonlocal term)} = 0.
\ea
Here the nonlocal term depends on the $AdS$ curvature $\ell$ and the brane intrinsic
curvature $H$, and is given by the time derivatives of the regular part $G^N$ of the
$AdS$ bulk Green function $G^N_{bulk}$ of the bulk wave equation:
\ba
\mbox{(nonlocal term)} & = & -{\kappa^2\dot{\phi}^2\over 6H}\int_0^{+\infty}
ds \left[\ddot{G}^N(s)+H\dot{G}^N(s)+{p^2\over
        a^2}G^N(s)\right]\dot{Q}(\tau-s).\quad
\ea
The local terms do not depend on the curvature, which means that the regular
part $G^N(\tau,\tau,0,0)$ of the bulk Green function is equal to the Minkowski propagator at the origin in
(\ref{qqq:geqscal}). Since the
Neumann form of the Minkowski propagator  $G^N_{Mink}(\tau,\tau',\xi,\xi')$ is
given by
\ba
G^N_{Mink}(\tau,\tau',\xi,\xi') &  = & G_{Mink}(\tau,\tau',\xi,\xi')+G_{Mink}(\tau,\tau',\xi,-\xi')
\ea
where
\ba
G_{Mink}(\tau,\tau',\xi,\xi') & = &
\theta\left((\tau-\tau')^2-(\xi-\xi')^2\right)J_0\left(p\sqrt{(\tau-\tau')^2-(\xi-\xi')^2}\right),
\ea
one has $G^N(\tau,\tau,0,0) = G^N_{Mink}(\tau,\tau,0,0) = 1$ and $\dot{G}^N(\tau,\tau,0,0) =
\dot{G}^N_{Mink}(\tau,\tau,0,0) = 0$ so that the effective equation of motion
for the inflaton reduces to
\ba
\left[1-{\kappa^2\dot{\phi}^2\over
    6H}\right]\ddot{Q}+\left[3H-{\kappa^2\dot{\phi}^2\over
    6H}H\right]\dot{Q}+{p^2\over a^2}Q+\mbox{(nonlocal term)} & = & 0.
\ea
We may renormalize the kinetic term to one by dividing this equation by the
coefficient in front of the kinetic term to obtain
\ba\label{qqq:effeominf}
\ddot{Q}+\left[3H+\Gamma(\tau)\right]\dot{Q}+\left({p^2\over a^2}+\Lambda(\tau)\right)Q+\mbox{ (nonlocal term)}+\mathcal{O}(\kappa^4\dot{\phi}^4) & = & 0.
\ea
From (\ref{qqq:effeominf}) we observe that the interaction of the inflaton with the bulk gravitons lead to
\emph{local} dissipation into the extra dimension through the local friction term appearing in the
effective equation as the first
time derivative $\Gamma(\tau)\dot{Q}$. There is in addition a phase shift
given by $\Lambda(\tau)Q$. We find that the local dissipation rate of the inflaton perturbation due to the extra
dimension is:
\ba
\Gamma(\tau) & = & {\kappa^2\dot{\phi}^2\over 3} \cr
             & = & \mathcal{O}(1){\dot{H}\over H^2}{H\ell\over\sqrt{1+H^2\ell^2}}H
\ea
and the local phase shift is
\ba
\Lambda(\tau) & = & \mathcal{O}(1){\dot{H}\over
  H^2}{H\ell\over\sqrt{1+H^2\ell^2}}{p^2\over a^2}.
\ea
We observe that the dissipation term $\Gamma\dot{Q}$ dominates the phase
correction $\Lambda Q$ at superhorizon scales and may be comparable at high-energy ($H\ell\gg 1$) to the slow-roll corrections to standard
inflation.
If inflation takes place in a regime where the bulk curvature radius
is much larger than the Hubble radius ($H\ell \gg 1$), then the local dissipation
rate behaves as
\ba
\Gamma(\tau) & \stackrel{H\ell \gg 1}{\sim} & \mathcal{O}(1){\dot{H}\over H^2}H.
\ea
In a quasi-four dimensional regime ($H\ell \ll 1$) the local dissipation rate
behaves as
\ba
\Gamma(\tau) & \stackrel{H\ell \ll 1}{\sim} & \mathcal{O}(1){\dot{H}\over H^2}(H\ell)H.
\ea
The dissipation rate is suppressed in a quasi-four dimensional regime by the
factor $\left(H\ell\right)$. The scalar field on the brane dissipates
linearly with the slow-roll parameter at any scale.

As we will see for tensor perturbations, the brane-bulk
interaction between the graviton bound state (zero mode) and the continuum of
bulk gravitons will be purely nonlocal since the brane degree of freedom here
is not confined but localized near the brane. In case of purely
nonlocal interaction we need to consider the bulk background geometry, which
is the object of the next section.

\section{Background metric}\label{sec:metric}

A flat (3+1)-dimensional homogeneous and isotropic Friedman-Robertson-Walker universe with
an arbitrary expansion history, characterized by the scale factor
$a(\tau )$ as a function of the proper time $\tau $, can be embedded in a wedge
of $AdS^5$ of curvature radius $\ell $,
described in terms of the
static
bulk coordinates (Poincar\'e coordinates)
\ba
 ds^2 & = & \ell^2 {-dt^2+d\mathbf{x}_{3}^2+dz^2 \over z^2},
\ea
by means of the following explicit embedding
\ba
z_b(\tau ) =
e^{-\int H(\tau)d\tau},\qquad
t_b(\tau )=
\int {d\tau\over\ell}\sqrt{\ell^2\dot{z_b}^2(\tau)+z_b^2(\tau)},
\ea
where $H(\tau)$ is the Hubble factor on the brane.
We may explicitly construct Gaussian normal coordinates by computing
the geodesic curve in the $zt$-plane normal to the braneworld trajectory
at proper time $\tau $, traveling at a proper distance
$\xi$ away from the brane. We thus obtain the following mapping from the Gaussian normal coordinates
to the Poincar\'e coordinates:
\ba
z(\xi,\tau) & = & \frac{e^{-\int d\tau ~H(\tau )}}
{ \cosh(\xi /\ell)
-\sqrt{1+\ell^2H^2}\sinh(\xi/ \ell)},\cr
t(\xi,\tau) & = &
\int{d\tau\over\ell}\left[e^{-\int d\tau ~H(\tau )}\sqrt{1+\ell^2H^2}\right]-{\ell He^{-\int
    d\tau ~H(\tau )}\sinh(\xi/ \ell )\over
\cosh(\xi/ \ell)-\sqrt{1+\ell^2H^2(\tau )}\sinh(\xi /\ell)},
\ea
from which follows the line element

\ba\label{eqnarray:exactmetric} ds^2 & = &
-\left(\sqrt{1+\ell^2
H^2}\sinh(\xi/\ell)-\cosh(\xi/\ell)+{\ell^2\dot{H}\over\sqrt{1+\ell^2H^2}}\sinh(\xi/\ell)\right)^2
d\tau^2\cr & &
+\left(\sqrt{1+\ell^2H^2}\sinh(\xi/\ell)-\cosh(\xi/\ell)\right)^2
e^{2\int H(\tau)d\tau} d\mathbf{x}_{3}^2\cr & & +d\xi^2. \ea
In these coordinates the brane is stationary with respect to the bulk, and $\xi$ measures
the proper distance from the brane.

\begin{figure}[htbp]
\setlength{\unitlength}{1cm}
\begin{center}
\begin{minipage}[t]{5.cm}
\begin{picture}(5.5,5.5)
\centerline {\hbox{\psfig{file=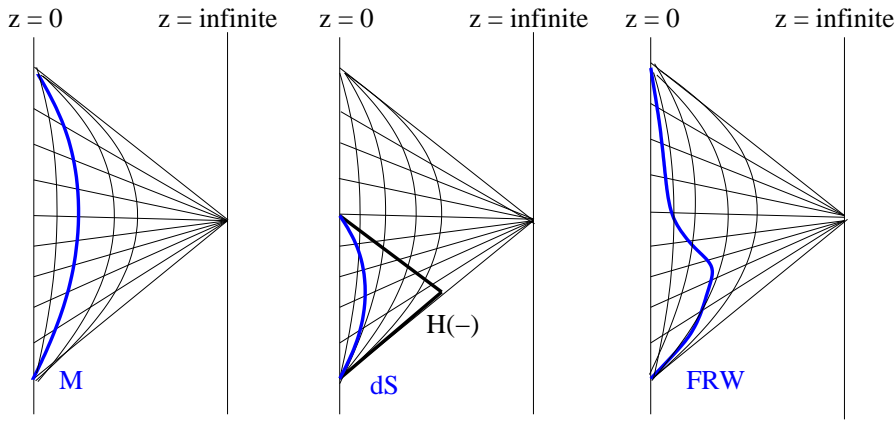,height=5.5cm}}}
\end{picture}\par
\end{minipage}
\hfill
\end{center}
\caption{\small{{\bf Carter-Penrose diagrams for various braneworld cosmologies embedded into
  $AdS$.} The heavy line indicates the trajectory of the timelike brane. The
  horizontal direction represents the extra dimension. The past Cauchy horizon
  is noted $H(-)$. From the left to the right: Minkowski braneworld, de Sitter
braneworld, Friedmann-Robertson-Walker braneworld. Panel on the right
describes an universe whose expansion initially inflates and then slows down.}} \label{fig:penrose}
\end{figure}

Despite being convenient
for describing a
brane with an arbitrary expansion history, Gaussian normal
coordinates suffer from a number of drawbacks. In particular this coordinate description can
break down whenever the spatial geodesics normal to the brane develop caustics,
either by focusing in time at $\xi_h  = \ell{\coth}^{-1}\left(\sqrt{1+\ell^2
    H^2}+\ell^2\dot{H}/\sqrt{1+\ell^2H^2}\right)$, or in the transverse
spatial dimensions at $\xi_s =  \ell{\coth}^{-1}\left(\sqrt{1+\ell^2
    H^2}\right)$.
Even though these singularities may give the illusion of a horizon, the bulk
space is regular there and can be extended beyond by using another set of coordinates.
Let us consider the initial value problem in these coordinates by looking at the Carter-Penrose diagram for Randall-Sundrum universes (Fig. \ref{fig:penrose}).
We specify initial data on a surface of constant Gaussian normal time, in the
bulk as well as on the brane. Suppose that we limit our ambition to predict
what happens on the brane in the future. The evolution of a dust-dominated
braneworld universe, which was initially inflating in the past, may be
causally affected in the future by unknown information coming from outside the
initial past Cauchy horizon, because the size of the bulk horizon has
increased during the deceleration of the expansion.

The complicated arbitrary motion of the brane into the bulk breaks the time-translation symmetry of
the bulk, such that the metric components in eqn. (\ref{eqnarray:exactmetric}) exhibit
a non-separable form. The evolution of the bulk perturbations in the
background metric (\ref{eqnarray:exactmetric}) is not amenable to analytic
methods without some form of approximation. Consequently, we will modify the
metric retaining only its most important features and assume that the bulk
solution is quasi-separable. From the point of view
of an observer on the brane, most of the action takes place within a small
number of $AdS$ decay length (or apparent decay length) of the brane, where
virtually all the bulk four-volume is concentrated.  The bulk
modes bound to the brane have most of their support localized there, therefore
a mediocre approximation of the metric far from the brane is only likely to
provide a poor approximation of the tail of the wave function, where there is
almost no probability. We may also hope that any quanta that escapes from the
brane, due to non-adiabatic effects in the expansion history, do not come back
through reflection or diffraction. Consequently, in this case an observer on
the brane will care little about how exactly such quanta escape, which will
depend on the metric away from the brane. In the WKB approximation, most of
the quanta become classical within a short distance from the brane. In that
sense we may approximate the background geometry (\ref{eqnarray:exactmetric})
by the geometry near the brane ($\xi \ll \ell$), by means of the following line element
\ba\label{eqnarray:approxmetric} ds^2 \approx
d\xi^2-e^{-2\alpha_1(\tau){\vert\xi\vert\over
\ell}}d\tau^2+e^{-2\alpha_2(\tau){\vert\xi\vert\over
\ell}}a^2(\tau)d\mathbf{x}_{3}^2, \ea where we have $Z_2$-symmetrized
about $\xi$ and where the time dependent warp factors and the scale factor are respectively given by \ba \alpha_1(\tau) & = &
\sqrt{1+\ell^2H^2}+{\ell^2\dot{H}\over\sqrt{1+\ell^2H^2}},\cr
\alpha_2(\tau) & = & \sqrt{1+\ell^2H^2},\cr a(\tau) & = & e^{\int
H(\tau) d\tau}. \ea
In this approximation (\ref{eqnarray:approxmetric}) the coordinate
singularities are also removed to infinity.

\section{Tensor perturbations in the approximate background}\label{sec:eom}

We now study the simplest case of tensor perturbations, which evolve
independently of the matter content on the brane. We use the approximate
background geometry (\ref{eqnarray:approxmetric}), which is accurate near the
brane. Each polarization of the tensor perturbations is described by a
minimally coupled massless scalar field  $\Psi$, as discussed in section \ref{sec:bcp}. We take the Fourier transform in the
three transverse spatial directions and separately evolve each Fourier mode.

\subsection{The plateau potential}\label{subsec:spectrum}

\begin{figure}[htbp]
\setlength{\unitlength}{1cm}
\begin{center}
\begin{minipage}[t]{4.cm}
\begin{picture}(4.5,4.5)
\centerline {\hbox{\psfig{file=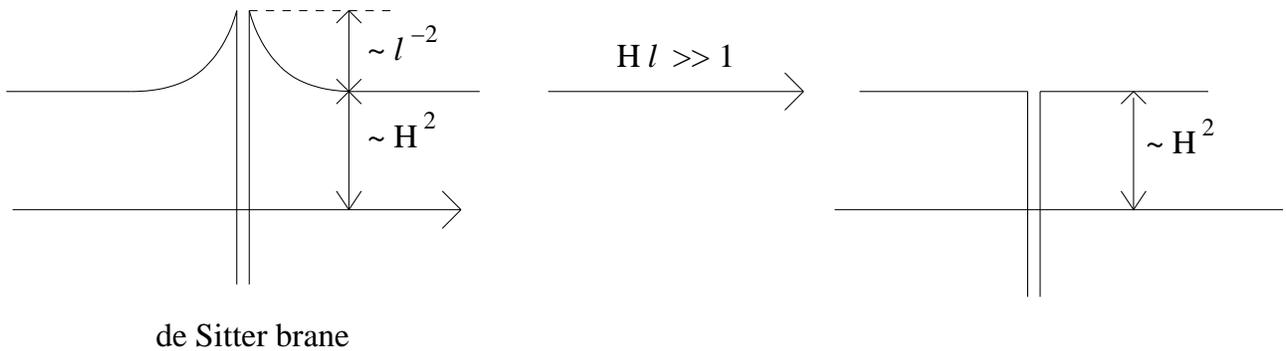,height=4.5cm}}}
\end{picture}\par
\end{minipage}
\hfill
\end{center}
\caption{\small{{\bf Approximation of the volcano potential by the plateau potential.}} } \label{fig:volcanolim}
\end{figure}

The equation of motion of the massless scalar field $\Psi$ in the background
metric (\ref{eqnarray:approxmetric}) is  \ba\label{eqnarray:firsteom} & &
\Biggl[-\partial_\tau^2-\left(3{\dot{a}\over
a}+\left({\dot{\alpha_1}\over\ell}-3{\dot{\alpha_2}\over\ell}\right)\vert\xi\vert\right)\partial_\tau+e^{-2\alpha_1{\vert\xi\vert\over
\ell}}\left(\partial_\xi^2-sgn(\xi)\left({\alpha_1\over\ell}+{3\alpha_2\over\ell}\right)\partial_\xi\right)\cr
& &-{p^2\over a^2}e^{-2\vert\xi\vert\left({\alpha_1\over
\ell}-{\alpha_2\over \ell}\right)}\Biggr]\Psi =  0,
\ea
where $\mathbf{p}$ is the momentum
in the three transverse dimensions. From this equation in the interval $-\infty < \xi <+\infty$, we may extract the homogeneous Neumann boundary
condition (\ref{eqnarray:bct}) at $\xi = 0$ for the even modes. After rescaling the scalar field as \ba\label{eqnarray:recale} \Phi(\tau,\xi) =
e^{-\left(\alpha_1(\tau)+3\alpha_2(\tau)\right){\vert\xi\vert\over
2\ell}}a(\tau)^{3\over
2}\Psi(\tau,\xi), \ea
the equation of motion becomes
\ba\label{eqnarray:rescaledeom} & &
\Biggl[-\partial_\tau^2+\mathcal{O}\left(\vert\xi\vert\right)\partial_\tau+\left({9\over 4}H^2+{3\over
2}\dot{H}-{p^2\over a^2(\tau)}e^{-2\vert\xi\vert\left({\alpha_1\over
\ell}-{\alpha_2\over \ell}\right)}+\mathcal{O}\left(\vert\xi\vert,\vert\xi\vert^2\right)\right)\cr
& &+e^{-2\alpha_1{\vert\xi\vert\over
\ell}}\left(\partial_\xi^2+\delta(\xi)\left({\alpha_1\over\ell}+{3\alpha_2\over\ell}\right)-{1\over 4}\left({\alpha_1\over\ell}+{3\alpha_2\over\ell}\right)^2\right)\Biggr]\Phi =  0.
\ea
The equation of motion is not separable yet. However, because the bound state
is localized near the brane, an adequate approximation to the evolution of the
bound state may be obtained by retaining only the leading behavior of the
coefficients in $\xi$ about $\xi = 0$, thus simplifying the equation. Doubtless this is a poor approximation for the tails of the bound state but we argue that the tails contribute negligibly. Consequently we set
\ba
e^{-2\alpha_1{\vert\xi\vert\over
\ell}} \approx 1, \qquad  e^{-2\vert\xi\vert\left({\alpha_1\over
\ell}-{\alpha_2\over \ell}\right)} \approx 1, \qquad
\mathcal{O}\left(\vert\xi\vert\right),\mathcal{O}\left(\vert\xi\vert^2\right)
\approx 0,
\ea
reducing eqn. (\ref{eqnarray:rescaledeom}) to
\ba\label{eqnarray:schro}
-\partial_\xi^2\Phi+V(\xi,\tau)\Phi =
-\partial_\tau^2\Phi+\left[{9\over 4}H^2+{3\over
2}\dot{H}-{p^2\over a^2(\tau)}\right]\Phi \ea with
the effective plateau potential \ba\label{eqnarray:potH} V(\xi,\tau) & = &
-V_0(\tau)\delta(\xi)+{V_0(\tau)^2\over 4}\ea
where
\ba
V_0(\tau)  & = & {1\over\ell}\left(4\sqrt{1+\ell^2H^2}+{\ell^2\dot{H}\over\sqrt{1+\ell^2H^2}}\right).
\ea
The corresponding boundary condition is obtained by integrating the
$Z_2$-symmetric equation (\ref{eqnarray:schro}). Equivalently, we may impose the boundary condition separately
\ba \left[\partial_\xi+{V_0(\tau)\over
    2}\right]\Phi\vert_{\xi = 0^+} = 0
\ea
at $\xi = 0$ and restrict the domain of (\ref{eqnarray:schro}) to $\xi > 0$,
thus removing the $\delta$-function. Under this approximation, the crater of
the usual volcano potential \cite{rs,lan} is faithfully rendered but the
landscape around the summit remains at constant elevation, as indicated in
Fig. \ref{fig:volcanolim} right.


When the brane has a pure de Sitter geometry, the linearized Einstein equations
in $AdS$ for the tensor perturbations may be reduced to a
"Schr\"{o}dinger-like" equation for a scalar field in the conformal bulk
coordinates once the field has been rescaled. The exact expression of the
effective potential has been calculated in \cite{lan}. The potential has the
usual volcano shape with a decreasing barrier potential resulting from the
curvature of the $AdS$ bulk (Fig. \ref{fig:volcanolim} left). The
height of the decreasing barrier potential is $\mathcal{O}(\ell^{-2})$ and the
energy gap between the zero mode and the continuous modes is
$\mathcal{O}(H^2)$.Thus the plateau potential approximation we presented
previously  consists in neglecting the inhomogeneity of the bulk and
consequently the diffraction of the bulk gravitons by the curved bulk. This
approximation is legitimized in the cosmological regime where the curvature
radius of the bulk is much larger than the Hubble horizon,
$H\ell\gg 1$. This is because in this regime the energy range of the barrier
potential is of order $\ell^{-2}$ and is much smaller than the Hubble
energy scale $H^2$ of the gap. In this regime the volcano potential looks like the
plateau potential (Fig. \ref{fig:volcanolim}),
in which case the bulk inhomogeneity may effectively be neglected.
In the regime where the curvature radius of the extra dimension is much
smaller than the Hubble horizon,
$H\ell \leq 1$, our approximation fails because the energy range
of the inhomogeneity of the bulk is now non negligible. Thus in that regime the plateau potential would only give some rough lower and upper bounds of the true volcano potential.

\subsection{Decay of the bound state for an inflating brane}

The potential in (\ref{eqnarray:potH}) has exactly one bound zero mode
and a continuum of free modes whose energy eigenvalues begin above the plateau.
We expand the wave function in normal modes
 \ba\label{eqnarray:modekg} \Phi(\tau,\xi) = c_b(\tau)\phi_b(\xi;\tau)+\int_0^{+\infty} dk
c_k(\tau)\phi_k(\xi;\tau) \ea where $b$ labels the bound mode and $k$ labels the continuum modes.  The mode functions $\phi_n(\xi;\tau)$ ($n = b,k$) satisfy \ba\label{eqnarray:eqproprekg}
\left(-\partial^2_\xi-V_0(\tau)\delta(\xi)+{V_0(\tau)^2\over
4}\right)\phi_n(\xi;\tau) = m_n^2(\tau)\phi_n(\xi;\tau). \ea
The $Z_2$-symmetric spectrum of (\ref{eqnarray:eqproprekg}) consists of a
single bound state with $m_b^2(\tau) = 0$,
\ba\label{eqnarray:zeromode} \phi_{b}(\xi;\tau) = N_b(\tau)
e^{-{V_0(\tau)\over 2}\vert\xi\vert},~~~~N_b(\tau) = \sqrt{V_0(\tau)/ 2} \ea
and a continuum of free bulk states $k$ with $m_k^2(\tau) =
k^2+V_0^2(\tau)/ 4$,
\ba\label{eqnarray:kmode} \phi_{k}(\xi;\tau) =
N_k(\tau)\left[\cos\left(k\xi\right)-{V_0(\tau)\over
2k}\sin\left(k\vert\xi\vert\right)\right],~~~~N_k(\tau) =
{k\over\sqrt{\pi}\sqrt{k^2+{V_0^2(\tau)\over 4}}}. \ea
A gap of $V_0(\tau)^2/ 4$ separates the bound state from the
continuum.

\begin{figure}[htbp]
\setlength{\unitlength}{1cm}
\begin{center}
\begin{minipage}[t]{4cm}
\begin{picture}(5.,5.)
\centerline {\hbox{\psfig{file=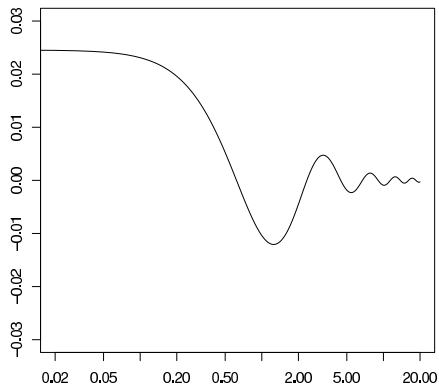,height=5.cm}}}
\end{picture}\par
\end{minipage}
\hfill
\end{center}
\caption{\small{{\bf Plot of the bulk interaction kernel $K(s)$ with respect
to time $s$, at superhorizon scales $p << aH$ and in the limit $H\ell\gg 1$}.
Here $H = 1$.}}\label{fig:kernel}
\end{figure}

We consider the universe during an inflationary epoch, when the geometry of the brane is quasi-de Sitter and the expansion is \emph{adiabatic} in the sense that $\dot{H} \ll H^2$. In this regime the brane-bulk interaction may be computed perturbatively. The only tensor degree of freedom on the brane is the discrete bound mode. We may compute the bound state decay rate due to the inflating motion of the brane into the bulk. Under the above approximation,
\ba\label{eqnarray:potadiab}
V_0(\tau) & \approx & 4\sqrt{1+\ell^2 H^2(\tau)}/\ell, \qquad \dot{V_0}
\approx 4\dot{H}{H\ell\over\sqrt{1+\left(H\ell\right)^2}}.
\ea
The equation for the time dependent mode expansion coefficients
$c_n(\tau)$ ($n = b,k$) in the approximate background described by eqn. (\ref{eqnarray:approxmetric}) is
\ba\label{eqnarray:unkgtau}
\ddot{c}_n+\left[{p^2\over a^2(\tau)}-{9\over 4}H^2+m_n^2(\tau)\right] c_n & \approx &
-2\dot{V}_0\left\langle\phi_n\Bigg\vert{\partial\phi_n\over\partial
  V_0}\right\rangle\dot{c}_n-\sum_{m\neq n}2\dot{V}_0\left\langle\phi_n\Bigg\vert{\partial\phi_m\over\partial
  V_0}\right\rangle\dot{c}_m\cr
& & +\mbox{ (higher order terms) }
\ea
where the higher order terms include two time derivatives acting on the potential
(e.g. $\ddot{H}, \dot{H}^2$). The terms $\left\langle\phi_n\vert\partial_{V_0}\phi_n\right\rangle$ vanish since the eigenmodes are
real. The terms $\left\langle\phi_k\vert\partial_{V_0}\phi_{k'}\right\rangle$ connecting two continuum states are also zero because of orthogonality\footnote{The orthogonality is the result of the simplified shape of the plateau potential.}. The only nonzero matrix elements connect the bound mode and a continuum mode, \ba
\left\langle\phi_{k}\Bigg\vert{\partial\phi_{b}\over\partial V_0}\right\rangle =
2N_k^{*}N_b\int_0^{+\infty}d\xi\left[\cos\left(k
\xi\right)-{V_0(\tau)\over
2k}\sin\left(k\xi\right)\right]\left(-{\xi\over
  2}\right)e^{-{V_0(\tau)\over 2}\xi}
 =  {N_k^{*}(\tau)N_b(\tau)\over
  k^2+{V_0^2(\tau)\over 4}}.
\ea
Equation (\ref{eqnarray:unkgtau}) reduces to
\ba
\ddot{c}_k+\Omega_k^2(\tau)c_k & \approx & -\gamma_k(\tau)\dot{c}_b(\tau)+\mathcal{O}\left(\dot{V_0}^2,\ddot{V_0}\right),\label{eqnarray:system2} \\
\ddot{c}_b+\Omega_b^2(\tau)c_b & \approx & \int dk \gamma_k(\tau)\dot{c}_k(\tau)+\mathcal{O}\left(\dot{V_0}^2,\ddot{V_0}\right),\label{eqnarray:system}
\ea
coupling the bound state to the continuum states with $k > 0$. In the
adiabatic approximation the first order coupling factor $\gamma_k(\tau)$ and
the frequencies $\Omega_b$, $\Omega_k$ are
\ba
\gamma_k(\tau) & = &
\dot{V_0}(\tau)\sqrt{2V_0(\tau)\over\pi}{k\over
  \left(k^2+{V_0^2(\tau)\over 4}\right)^{3/2}},\cr
\Omega_b^2(\tau) & \approx
&{p^2\over a^2(\tau)}-{9\over
    4}H^2(\tau),\cr
\Omega_k^2(\tau) & \approx & {p^2\over a^2(\tau)}-{9\over
    4}H^2(\tau)+k^2+{V_0^2(\tau)\over 4}.
\ea
The time dependent coupling factor in (\ref{eqnarray:system2}) leads to
transitions between the bound mode and the continuous modes, because the
zeroth order bare modes act as a source for the modes at next
order. As soon as the acceleration of the brane changes in the bulk, through the
derivative of the Hubble factor, $\dot{H}(\tau)$ (or equivalently
$\dot{V_0}(\tau)$), the \emph{in} mode acts as a source through the coupling
factor and generates different modes at first
order. Consequently transitions from the bound mode to the continuum occur at first
order, and the first order continuous modes act then themselves as a source for
the bound mode, as a back scattering effect at second order. The brane bound mode and the bulk continuous modes are related according to
\ba
c_k(\tau)  & = & \int_{-\infty}^\tau d\tau' G^k_{bulk}(\tau,\tau')\left(-\gamma_k(\tau')\right)\dot{c}_b(\tau'),\cr
c_b(\tau) & = & \int_{-\infty}^\tau d\tau' G_{brane}(\tau,\tau')\int dk \gamma_k(\tau')\dot{c}_k(\tau'),
\ea
where $G^k_{bulk}(\tau,\tau')$ and $G_{brane}(\tau,\tau')$ are the bare
retarded Green functions satisfying
\ba\label{eqnarray:related}
\left(\partial_\tau^2 + \Omega_k^2(\tau)\right)G^k_{bulk}\left(\tau,\tau'\right) & = & \delta\left(\tau-\tau'\right),\cr
\left(\partial_\tau^2 + \Omega_b^2(\tau)\right)G_{brane}\left(\tau,\tau'\right) & = & \delta\left(\tau-\tau'\right).
\ea
We may express the retarded bulk Green function of a mode $k$ in the WKB approximation
\ba
G^k_{bulk}(\tau,\tau') & \approx & \theta(\tau-\tau'){\sin\left(\int_{\tau'}^\tau
    \Omega_k(\tau)d\tau\right)\over \sqrt{\Omega_k(\tau)\Omega_k(\tau')}}.
\ea
This WKB approximation is reasonable because that bulk eigenmodes oscillate at
all times. There is no turning point because $\Omega_k^2 \approx
p^2/a^2-9H^2/4+k^2+V_0^2/4 > 0$ always holds. The condition
$\dot{\Omega}_k \ll \Omega_k^2$ is verified for both subhorizon modes, $p \gg aH$, and superhorizon modes, $p \ll aH$.
The effective brane propagator with the interaction with the bulk modes of freedom taken into account is obtained by summing the infinite geometric series
\ba\label{eqnarray:effgreentensor}
\hat{G}_{brane} & = &  G_{brane} + G_{brane}\left[\int dk \gamma_k D_\tau G^k_{bulk}\left(-\gamma_k\right)D_\tau\right] G_{brane}\cr
& + & G_{brane}\left[\int dk \gamma_k D_\tau G^k_{bulk}\left(-\gamma_k\right)D_\tau\right] G_{brane}\left[\int dk \gamma_k D_\tau G^k_{bulk}\left(-\gamma_k\right)D_\tau\right] G_{brane}
 + ...\cr
              & = & {1\over G^{-1}_{brane}+\int dk \gamma_k D_\tau G^k_{bulk}\gamma_k D_\tau}.
\ea
The four dimensional interaction between the brane modes due to the expansion of the universe is contained in $G^{-1}_{brane}$, whereas the interaction between the bound mode and the bulk continuum is contained in the bulk kernel
\ba\label{eqnarray:self}
K(\tau,\tau') & = & \int dk \gamma_k(\tau) \partial_\tau G^k_{bulk}(\tau,\tau')\gamma_k(\tau'),\cr
               & = & \theta(\tau-\tau')\int dk\gamma_k(\tau)\sqrt{{\Omega_k(\tau)\over \Omega_k(\tau')}}\cos\left(\int_{\tau'}^\tau
    \Omega_k(\tau)d\tau\right)\gamma_k(\tau'),
\ea
such that the brane bound mode is governed by the integro-differential equation
\ba\label{eqnarray:exacteq}
 &  & \hat{G}_{brane}^{-1}\circ c_b\cr
  & = & \partial_\tau^2c_b(\tau)+\Omega_b^2(\tau)c_b(\tau)+\int_{-\infty}^\tau
  d\tau' K(\tau,\tau')\partial_{\tau'}c_b(\tau') = 0
\ea
where the bulk interaction kernel $K(\tau,\tau')$ dresses the
bare field of the bound state. The possible dissipation of the bound state appears
 to be \emph{nonlocal} as viewed by an observer on the brane. This
feature arises because the bound state is not a purely local but a localized
brane degree of freedom with a typical extent in the bulk given by $\ell_{att}
= (V_0/2)^{-1}$. The existence of the bound state is intrinsically dependent
on the bulk curvature $\ell^{-1}$. This condition is incompatible with a local
interaction with the bulk. In the limit $H\ell \gg 1$ one has $V_0\approx 4H$, $\dot{V_0} = 4\dot{H}$. The bulk interaction kernel (\ref{eqnarray:self}) reduces in the limit $H\ell
\gg 1$ to
\ba\label{eqnarray:superself}
K(s\equiv \tau-\tau') & \stackrel{H\ell\gg 1, p\ll aH}{\approx} &
\theta(s)\dot{H}^2\int_0^\infty dk {128k^2 H\over
  \pi\left(k^2+4H^2\right)^3} \cos\left(\sqrt{k^2+{7\over
      4}H^2}~s\right)
\ea
at superhorizon scales,
and to
\ba\label{eqnarray:subself}
K(\tau,\tau') & \stackrel{H\ell\gg 1, p\gg aH}{\approx} &
\theta(\tau-\tau')\dot{H}^2\int_0^\infty dk {128k^2 H\over
  \pi\left(k^2+4H^2\right)^3} \left({k^2+{p^2\over a(\tau)^2}\over k^2+{p^2\over
    a(\tau')^2}}\right)^{1/4}\cos\left(\int_{\tau'}^\tau \sqrt{k^2+{p^2\over
    a(\bar{\tau})^2}}d\bar{\tau}\right)\cr
              & \approx & \mathcal{O}(1)H^2\left({\dot{H}\over
H^2}\right)^2e^{-{H\over 2}(\tau-\tau')}\cos\left({p\over H}\left(e^{-H\tau}-e^{-H \tau'}\right)\right)\theta(\tau-\tau')
\ea
at subhorizon scales, because the integral over $k$ is suppressed for $k \gg H$
in that case.
We used that $H$, $\dot{H}$ are fairly constant in time in the adiabatic
approximation. The superhorizon kernel (\ref{eqnarray:superself}) is
plotted in Fig. \ref{fig:kernel}.
We observe that the kernel is nonlocal, typically varying with the Hubble
time scale $H^{-1}$. Thus the bulk interaction contains a memory.
Consequently a local approximation would be possible only if the bound mode varied typically
with a time scale larger than the Hubble time.

During inflation, the Hubble radius of the universe remains fairly constant,
whereas the physical wavelength of a mode increases exponentially due to the
expansion. Early during inflation a mode of a given wave number is initially
subhorizon so that $p\gg a(\tau)H$. As the universe expands, the mode crosses
the horizon to become superhorizon. Whereas modes
oscillate at subhorizon scales, they become frozen in after they exit the Hubble
horizon. An observer is sensitive to the superhorizon modes on the brane which
 are the relevant modes to present cosmological observations and constitute the ``initial conditions'' of the universe at the end of inflation.
 In that sense we will consider
mostly the superhorizon modes, or late-time asymptotics of the modes:
\ba
p\ll a(\tau)H.
\ea
The evolution of the modes at superhorizon scales taking into
account the bulk interaction through the integral kernel is
nonlocal. Consequently the evolution (and the possible dissipation) of superhorizon modes depends on
their past evolution at subhorizon scales through the interaction of the
subhorizon modes with the bulk.
At superhorizon scales the bare frequency of the bound mode is imaginary
\ba
\Omega_b^2(\tau) = -{9\over 4}H^2(\tau).
\ea
The bare bound mode coefficient $c_b^{(0)}(\tau)$ excluding the interaction with the bulk is a
superposition of a growing mode and a decreasing mode
\ba\label{qqq:cbare}
c_b^{(0)}(\tau)  & = & A_+e^{+{3\over 2}\int H(\tau)d\tau} +A_-e^{-{3\over
    2}\int H(\tau)d\tau},\cr
                 & \approx & A_+e^{+{3\over 2}H\tau} +A_-e^{-{3\over
    2}H\tau},
\ea
 at superhorizon scales and in the WKB approximation. Since we have rescaled
 the fields in (\ref{eqnarray:recale}) by the factor $a^{3/2}(\tau)$, the true
 bound mode is actually a superposition
 of a constant mode and a decreasing mode. The bare
 solutions (\ref{qqq:cbare}) correspond also to the superhorizon asymptotics of the four
 dimensional solution given by the Hankel function in the limit of a
 pure de Sitter four dimensional geometry,
\ba
H^{(1),(2)}_{3/2}\left(-p\eta\equiv {p\over a(\tau)H} \right)& \stackrel{p/aH \ll
  1}{\longrightarrow}& \sim A_+a^{3/2}(\tau)+A_-a^{-3/2}(\tau),\cr
                     & \stackrel{p/aH \gg
  1}{\longrightarrow}& \sim e^{\pm i {p\over a(\tau)H}}
\ea
where $\eta$ is the conformal proper time and $a(\tau) = e^{H\tau}$.
 The decreasing mode is unobservable and has little
 cosmological interest for the formation of the large scale structures in the
 universe. In the following we consider only the dominant solution composed of the growing
 mode.

 We look now for the growing mode of the dressed
 bound state coefficient with the interaction with the bulk taken into account
 by using the superhorizon Ansatz
\ba\label{eqnarray:ansatz}
c_b(\tau) & = & \exp[+\gamma \tau]
\ea
where
\ba
\gamma & = & {3\over 2}H+\Delta\gamma,\qquad \Delta\gamma\ll H.
\ea
Since the bulk interaction kernel is nonlocal, we should take care when
inserting this superhorizon Ansatz into the integral kernel in
eqn. (\ref{eqnarray:exacteq}) because the rigorous complete integration should take into
account the contribution of the subhorizon regime, where the mode has a
 form different from the
superhorizon Anstaz (\ref{eqnarray:ansatz}). However, we conjecture that, when
observing the dressed superhorizon solution today, the integrated effects of
the subhorizon modes in the brane-bulk interaction are subdominant compared to the
contribution of the superhorizon modes. This is because the amplitude
of subhorizon modes is suppressed compared to the amplitude of
superhorizon modes. The choice of the Ansatz (\ref{eqnarray:ansatz}) at any time for
solving the integro-differential equation (\ref{eqnarray:exacteq}) for the
bound state is equivalent to smooth the weak amplitude subhorizon oscillations
of the complete mode solution.
In that sense we consider the superhorizon Ansatz
(\ref{eqnarray:ansatz}) sufficient to solve the integro-differential equation
and compute the order of magnitude of the dissipation rate of the bound state
at superhorizon scales.

We now proceed to compute the attenuation factor of the growing bound mode in the regime $H\ell \gg 1$.
In this regime the curvature radius of the extra dimension is much
larger than the Hubble radius so that the dissipation rate of the bound state
into the extra dimension is expected to reach its maximal value. As we
discussed in section \ref{subsec:spectrum}, the plateau potential is a
suitable approximation in this regime for the study of the brane-bulk interaction. The
height of the plateau potential is then $V_0^2(\tau)/ 4  \approx
4H^2(\tau)$ and the dressed bound mode
coefficient evolves at superhorizon scales according to the equation
\ba\label{eqnarray:effeom2}
\ddot{c}_b(\tau) -{9\over 4}H^2(\tau)c_b(\tau)+\int_{0}^{+\infty}
  ds K(s)\dot{c}_b(\tau-s) = 0,
\ea
where $K(s)$ is the bulk interaction kernel taken in the superhorizon limit, eqn. (\ref{eqnarray:superself}).

Inserting the Ansatz (\ref{eqnarray:ansatz}) into (\ref{eqnarray:effeom2}) yields the algebraic
equation
\ba
\gamma^2-{9\over 4}H^2 & = & -{128\over\pi}\dot{H}^2H\int_0^\infty dk {k^2\gamma^2\over
  \left(k^2+4H^2\right)^3\left(k^2+{7\over 4}H^2+\gamma^2\right)},
\ea
which to linear order in $\Delta\gamma$ becomes
\ba
3H\Delta\gamma & \approx & -{128\over\pi}\dot{H}^2H\left[{9\over 4}H^2\int_0^\infty dk {k^2\over
      \left(k^2+4H^2\right)^4}\right]
\ea
such that
\ba
\Delta\gamma & \approx & -{3\over 2^{5}}H\left({\dot{H}\over
    H^2}\right)^2.
\ea

The minus sign in $\Delta\gamma$ induces attenuation of the growing
mode. The plateau potential approximation has the advantage of eliminating the
processes of diffraction or reflection in the bulk by smoothing the bulk
inhomogeneity, keeping only the dissipative processes due to the extra
dimension. Even if the bound state did not dissipate at subhorizon scales
and had only its frequency shifted, the bound mode
would be observed to dissipate classically in the extra dimension at superhorizon scales as
\ba
c_b(\tau) & = & e^{-\int\Gamma(\tau)d\tau}c_b^{(0)}(\tau)
\ea
where $c_b^{(0)}(\tau)$ is the bare growing mode. The dissipation
rate $\Gamma(\tau) \equiv -\Delta\gamma$ of the bound state at superhorizon scales is given by
\ba
\Gamma(\tau) & = & \mathcal{O}(1)\left({\dot{H}\over
    H^2}\right)^2 H.
\ea
The emitted bulk gravitons
oscillates at all times because their frequency remains above the plateau of
the potential.
We may also express the attenuation factor in terms of the four dimensional slow-roll inflation
parameter $\epsilon_H = -\dot{H}/H^2$, characterizing the adiabaticity of the
expansion of the universe, and the number $N$ of  e-folds as
\ba
e^{-\int\Gamma(\tau)d\tau} & = & e^{-\mathcal{O}(1)\epsilon_H^2N}.
\ea

The graviton bound state decays nonlocally and quadratically in the slow-roll
factor. The dissipation of the zero mode of the graviton during inflation is thus
subdominant compared to the dissipation of the inflaton perturbation (section \ref{sec:scalar2}).

\section{Scalar perturbations in the approximate background: adiabatic perfect
  fluid on the brane}\label{sec:scalar}

In this section we study the evolution of scalar metric perturbations in the
approximate background geometry (\ref{eqnarray:approxmetric}) accurate near the brane. These
perturbations couple to the adiabatic perturbations of a perfect fluid on the brane. They are
all described by the single master field  $\Omega$ as previously discussed
in section \ref{sec:bcp}. We take the Fourier transform in the
three  spatial transverse directions because of the homogeneity and the
isotropy, and separately evolve each Fourier mode. In the approximate
background geometry (\ref{eqnarray:approxmetric}) near the brane, the equation
of motion (\ref{eqnarray:eoms}) of the master field in the interval $ 0 < \xi
<+\infty$ is:
\ba\label{eqnarray:eomss} & &
\Biggl[-\partial_\tau^2+\left(3{\dot{a}\over
a}-\left({\dot{\alpha_1}\over\ell}+3{\dot{\alpha_2}\over\ell}\right)\xi\right)\partial_\tau+e^{-2\alpha_1{\xi\over
\ell}}\left(\partial_\xi^2-\left({\alpha_1\over\ell}-{3\alpha_2\over\ell}\right)\partial_\xi\right)\cr
& &-{p^2\over a^2}e^{-2\xi\left({\alpha_1\over
\ell}-{\alpha_2\over \ell}\right)}+{e^{-2\alpha_1{\xi\over
\ell}}\over\ell^2}\Biggr]\Omega =  0,
\ea
where $p$ is the transverse momentum. Similarly
to the tensor case in section \ref{sec:eom}, we reduce
the equation of motion (\ref{eqnarray:eomss}), rescaling
\ba\label{qqq:rescalemaster}
e^{-\left({\alpha_1-3\alpha_2\over\ell}\right){\xi\over
    2}}a(\tau)^{-3/2}\Omega & \rightarrow & \Omega,
\ea
to the equation
\ba\label{eqnarray:rescaledeoms} & &
\Biggl[-\partial_\tau^2+\mathcal{O}\left(\xi\right)\partial_\tau+\left({9\over 4}H^2-{3\over
2}\dot{H}-{p^2\over a^2(\tau)}e^{-2\xi\left({\alpha_1\over
\ell}-{\alpha_2\over \ell}\right)}+{e^{-2\alpha_1{\xi\over
\ell}}\over\ell^2}+\mathcal{O}\left(\xi,\xi^2\right)\right)\cr
& &+e^{-2\alpha_1{\xi\over
\ell}}\left(\partial_\xi^2-{1\over 4}\left({\alpha_1\over\ell}-{3\alpha_2\over\ell}\right)^2\right)\Biggr]\Phi =  0
\ea
which contains terms dependent on $\xi$ representing the inhomogeneity
of the $AdS$ bulk. The equation of motion is not separable yet. However, as in section
\ref{sec:eom} for the tensor case when we introduced the plateau potential, we
again simpliplify the dependence on $\xi$ retaining only the leading behavior
of the coefficients in $\xi$ about $\xi = 0$. We set
\ba
e^{-2\alpha_1{\xi\over
\ell}} \approx 1, \qquad  e^{-2\xi\left({\alpha_1\over
\ell}-{\alpha_2\over \ell}\right)} \approx 1, \qquad
\mathcal{O}\left(\xi\right),\mathcal{O}\left(\xi^2\right)
\approx 0,
\ea
reducing (\ref{eqnarray:rescaledeoms}) to
\ba\label{qqq:schroscalar}
\left[\partial_\tau^2-\partial_\xi^2+\bar{\omega}^2(\tau)+\lambda^2(\tau)\right]\Omega & = & 0
\ea
where
\ba
\lambda(\tau) & = &{\alpha_1-\alpha_2\over 2\ell}
= {\ell\dot{H}\over 2\sqrt{1+\ell^2H^2}} ~\leq 0,\label{eqnarray:v1}\cr
\bar{\omega}^2(\tau) & = & {p^2\over a^2}-{5\over 4}H^2+{1\over
2}\dot{H}.\label{eqnarray:omtilde}
\ea
The bulk curvature $\ell$ is contained in $\lambda$  and the intrinsic brane
curvature $H$ is contained in $\bar{\omega}$.
We highlight the ``wrong'' sign in the potential $\lambda(\tau)$ for
the scalar perturbations since $\dot{H}(\tau)<0$. The brane behaves as a repulsive potential
for the scalar perturbations so that there is no scalar gravitational bound state
near the brane as long as there is no coupling to matter on the brane.

The case of a perfect fluid on the brane (with equation of state $P_M =
w\rho_M$) gives rise to some difficulties in
computing the evolution of perturbations because the expansion of the brane
is no longer adiabatic. $\dot{H}/H^2$ is not small except when the
fluid is subdominant compared to the brane tension $\sigma$ (cosmological
constant). Here we explore the damping of the fluid perturbation on subhorizon
scales because in this regime the adiabatic approximation of the expansion
is valid. At subhorizon scales the geometry of the brane appears
quasi-Minkowskian.

The boundary conditions (\ref{eqnarray:sbc0}) linking the master field to the
fluid perturbations on the brane become in
the approximate background geometry (\ref{eqnarray:approxmetric}) near the brane
\ba\label{eqnarray:sbc}
\kappa^2 a\delta\rho & = & -3H\left(\dot{\Omega}'+{\alpha_1\over\ell}\dot{\Omega}\right)-{p^2\over
a^2}\left(\Omega'+{\alpha_2\over\ell}\Omega\right)\Bigr\vert_{\xi = 0},\cr
\kappa^2 a\delta
q & = & -\dot{\Omega}'-{\alpha_1\over\ell}\dot{\Omega}\Bigr\vert_{\xi = 0},\cr
\kappa^2 a\delta
P & = &
\left(\ddot{\Omega}'+{\alpha_2\over\ell}\ddot{\Omega}\right)+2H\left(\dot{\Omega}'+{\alpha_1\over\ell}\dot{\Omega}\right)+\left({\dot{\alpha_1}\over\ell}+2{\dot{\alpha_2}\over\ell}\right)\dot{\Omega}\cr
                 &   & +{\dot{\alpha_2}\over\ell
H}\left({1\over\ell^2}-{2\over 3}{p^2\over a^2}\right)\Omega-{\dot{\alpha_2}\over\ell
H}\left({\alpha_1\over\ell}-2{\alpha_2\over\ell}\right)\Omega'\Bigr\vert_{\xi = 0}.
\ea
It is possible to simplify the boundary conditions (\ref{eqnarray:sbc}) for
the scalar metric perturbations by expressing them in terms of the
gauge-independent matter perturbation $\left(\rho-\sigma\right)\Delta = \delta\rho-3H\delta
q$. One obtains the following boundary condition
\ba\label{eqnarray:bc}
\left(\Omega'+{\alpha_2(\tau)\over\ell}\Omega\right)\Bigr\vert_{\xi = 0} & = &
-\kappa^2{\left(\rho-\sigma\right) a^3\over p^2}\Delta(\tau).
\ea
which, in terms of the rescaled master field (\ref{qqq:rescalemaster}), becomes
\ba
\left(\partial_\xi+\lambda(\tau)\right)\Omega\vert_{\xi = 0} = -\kappa^2{\left(\rho-\sigma\right) a^{3/2}\over p^2}\Delta(\tau),
\ea
where $\lambda$ is given in (\ref{eqnarray:v1}).

The equation of motion for the matter perturbations on the brane can also be derived
from the boundary conditions (\ref{eqnarray:sbc}) by imposing an equation of state on the matter
perturbations. We consider adiabatic matter perturbations with
\ba
\delta P & = & c_s^2 \delta\rho,
\ea
where the sound speed of the fluid is $c_s^2 = \dot{P}/\dot{\rho}$. One obtains the
equation of motion for the matter perturbation $\Delta$, similar to the equation
obtained in
 \cite{scal},
\ba\label{eqnarray:eqm}
& \ddot{\Delta}+ & \left(8+3c_s^2-6\epsilon\right)H\dot{\Delta}+\left[{p^2c_s^2\over
  a^2}+\left(10+6c_s^2-14\epsilon+3\epsilon^2\right){\kappa^2\rho_M\over
  2\ell}+\left(5+3c_s^2-9\epsilon\right){\kappa^4\rho_M^2\over 12}\right]\Delta\cr
& = & \epsilon{p^4\over 3 a^5}a^{3/2}\Omega\vert_{\xi = 0},
\ea
where according to the braneworld dynamics \cite{bdel},\cite{deff}
\ba\label{qqq:bwdyn}
H^2 & = & {\kappa^2\over 3\ell}\rho_M\left(1+{\kappa^2\ell\over
    12}\rho_M\right) = -{1\over\ell^2}+{\kappa^4\over 36}\rho^2,\cr
\dot{\rho} & = & -3H(P+\rho) = -3H(1+w)\rho_M,\cr
 \epsilon & = & 1+w, \qquad \rho_M = \rho-\sigma, \qquad P_M = P+\sigma.
\ea
The equation of state of the fluid on the brane is $P_M = w\rho_M$. We
may also rescale the brane degree of freedom as $e^{(1/2)\int
  \left(8+3c_s^2-6\epsilon\right)H d\tau}\Delta \rightarrow \Delta$ to obtain
an oscillator equation. Therefore the equations coupling the brane scalar
perturbations to the bulk scalar perturbations may be summarized in the
generic form
 \ba
\ddot{\Omega}-\Omega''+\bar{\omega}^2(\tau)\Omega+\lambda^2(\tau)\Omega & = & 0, \cr
\Omega'\vert_{\xi = 0}+\lambda(\tau)\Omega\vert_{\xi = 0} & = &
-\gamma_1(\tau)\Delta,\cr
\ddot{\Delta}+\omega_0^2(\tau)\Delta & = & \gamma_2(\tau) \Omega\vert_{\xi =
0}\label{eqnarray:genm},
\ea
where $\lambda$ and $\bar{\omega}$ are given in (\ref{eqnarray:omtilde}) and
\ba\label{eqnarray:param}
\omega^2_0(\tau) & = & {p^2c_s^2\over
  a^2}+\left(10+6c_s^2-14\epsilon+3\epsilon^2\right){\kappa^2\rho_M\over
  2\ell}+\left(5+3c_s^2-9\epsilon\right){\kappa^4\rho_M^2\over 12}\cr
  & & -{1\over 4}\left(8+3c_s^2-6\epsilon\right)^2H^2-{1\over 2}\left(8+3c_s^2-6\epsilon\right)\dot{H}+9H^2\epsilon(\epsilon-c_s^2-1),\cr
\gamma_1(\tau) & = & \kappa^2{\rho_M a^{3/2}\over p^2}e^{-(1/2)\int \left(8+3c_s^2-6\epsilon\right)H d\tau},\cr
\gamma_2(\tau) & = & \epsilon{p^4\over 3 a^5}a^{3/2}e^{+(1/2)\int \left(8+3c_s^2-6\epsilon\right)H d\tau}.
\ea
The immediate observation is the absence of time derivative couplings for the
brane-bulk system in (\ref{eqnarray:genm}). Consequently the
dissipation of the brane fluid into the bulk can only be nonlocal despite the perfect
fluid being local (\emph{i.e.} truly localized on the brane). Say otherwise, no local friction term such as $\Gamma(\tau)\partial_\tau$
can appear in the effective equation on the brane, this means that the dissipation of the
brane fluid contains a
memory or a time delay. The nonlocal brane-bulk interaction comes
from the curvature effects.

\begin{figure}[htbp]
\setlength{\unitlength}{1cm}
\begin{center}
\begin{minipage}[t]{5.cm}
\begin{picture}(5.5,5.5)
\centerline {\hbox{\psfig{file=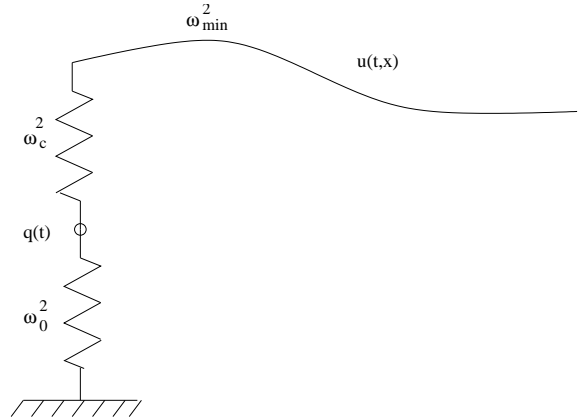,height=5.5cm}}}
\end{picture}\par
\end{minipage}
\hfill
\end{center}
\caption{\small{{\bf Mechanical model for braneworld scalar perturbations.} The string,
  representing the bulk gravitons, is coupled by an intermediate spring to a harmonic
oscillator representing the brane fluid.}} \label{fig:spring}
\end{figure}

The underlying physics can be illustrated using the following mechanical
model. Consider a string coupled to a harmonic oscillator at the boundary with
two springs as indicated in Fig. \ref{fig:spring}.
The equations of this model are similar to those obtained for the scalar
braneworld perturbations
\ba
\ddot{u}-u''+\omega^2_{min}(t) u & = & 0,\label{eqnarray:string} \cr
\gamma(t) u'\vert_{x = 0}-\omega^2_c(t) u\vert_{x = 0} & = &
-\omega^2_c(t) q(t),\label{eqnarray:link}\cr
\ddot{q}+\left(\omega_0^2(t)+\omega_c^2(t)\right) q & = & \omega^2_c(t) u\vert_{x = 0},\label{eqnarray:osc}
\ea
where $\gamma $ is the string tension expressed as a frequency, and
$\omega _0^2$ and $\omega _c^2$ are the spring constants of
the two springs expressed as frequencies squared. A toy model very similar to this mechanical system was considered in \cite{reso}.
 The intermediate spring between the mass and the string plays the role of a
buffer. In the absence of the buffer the coupling with the string would bring
a first derivative local term (friction term) in the effective equation for
the harmonic oscillator, indicating a local dissipation of the oscillator down the string.

With no uniform restoring force on the string,
the equations would be
\ba
\ddot u(x,t)-u^{\prime \prime }(x,t)&=&0,\cr
\gamma u^{\prime }(x=0,t)-{\omega _c}^2u(x=0,t)&=&-{\omega _c}^2q(t),\cr
\ddot q(t)+
({\omega _0}^2+{\omega _c}^2)q(t)-
{\omega _c}^2u(x=0,t)
&=&
f(t)
\ea
where
$f(t)$ is a forcing term. For coefficients not depending on time, the ans\"atze
$u(x,t)=u\cdot \exp [i\omega (x-t)],$
$q(t)=q\cdot \exp [-i\omega t],$ and
$f(t)=f\cdot \exp [-i\omega t]$
give
\ba
-\omega ^2q(\omega )+
({\omega _0}^2+{\omega _c}^2)q(\omega )-
\frac{{\omega _c}^4}{\omega _c^2-i\gamma \omega }q(\omega )=f(\omega ),
\ea
which may be rewritten as
\ba
\ddot q(t)+
({\omega _0}^2+{\omega _c}^2)q(t)+
\int _0^\infty ds~K(s)~q(t-s)
&=&
f(t)
\label{qqq:a}
\ea
where
\ba\label{qqq:kno}
K(t)&=&-\int _{-\infty }^{+\infty }
\frac{d\omega }{(2\pi )}
\frac{{\omega _c}^4~\exp [-i\omega t]}{{\omega _c}^2-i\omega \gamma }\cr
&=&-\frac{{\omega _c}^4}{\gamma }~\theta (t)~\exp [-{\omega _c}^2t/\gamma ]\cr
&=&-({\omega _c}^2\lambda )~\theta (t)~\exp [-\lambda t]
\ea
and $\lambda ={\omega _c}^2/\gamma .$ Consequently, eqn.~(\ref{qqq:a})
becomes
\ba
\ddot q(t)+
({\omega _0}^2+{\omega _c}^2)q(t)-
{\omega _c}^2\lambda
\int _0^\infty ds~\exp [-\lambda s]~q(t-s)
&=&
f(t).
\label{qqq:b}
\ea
In the limiting case where $\lambda \gg \omega _0, \omega _c,$ we may
approximate
\ba
\lambda\int _0^\infty ds~\exp [-\lambda s]~q(t-s)\approx q(t)-\lambda^{-1}\dot q(t),
\ea
so that eqn.~(\ref{qqq:b}) may be approximated as
\ba
\ddot q(t)+\gamma
\dot q(t)
+{\omega _0}^2q(t)=f(t)
\label{qqq:c}
\ea

\begin{figure}[htbp]
\setlength{\unitlength}{1cm}
\begin{center}
\begin{minipage}[t]{4.cm}
\begin{picture}(5.,5.)
\centerline {\hbox{\psfig{file=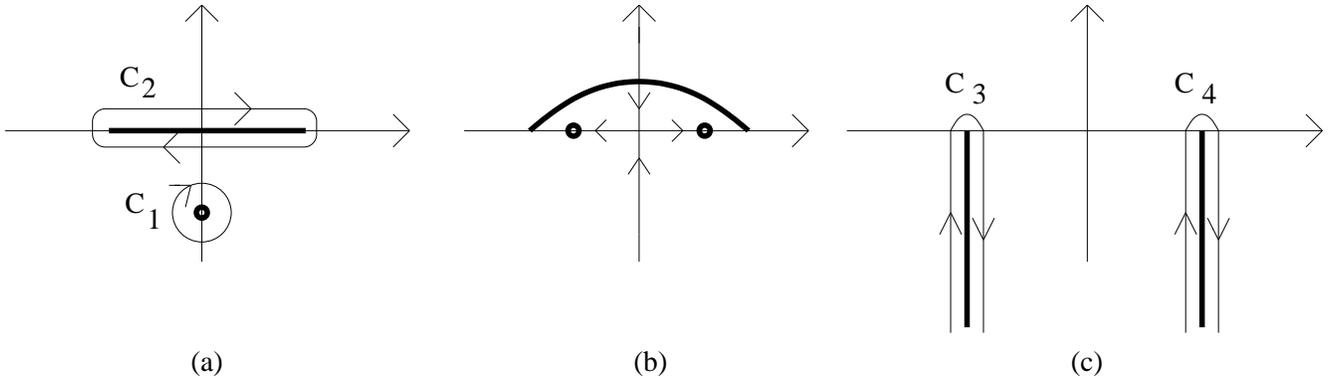,height=5cm}}}
\end{picture}\par
\end{minipage}
\hfill
\end{center}
\caption{\small{{\bf Analyticity properties of the kernel in the complex plane}. Panel (a):
if $\omega_{min} < \lambda=\omega_c^2/\gamma$ then there is an isolated pole
on the lower imaginary axis and a branch cut $[-\omega_{min},+\omega_{min}]$
on the real axis. Panel (b) and (c): if $\omega_{min} >
\lambda=\omega_c^2/\gamma$ then the pole moves to the real axis under the
branch cut as well as its mirror image coming from the second Riemann
sheet. We may deform the branch cut by pushing it to negative infinity (panel (c)).}  } \label{fig:pole}
\end{figure}

We next consider the wave equation for the string including
a restoring term, so that there is a minimum frequency $\omega _{min}$
for the propagating modes. One has
\ba
\ddot u-u^{\prime \prime }+{\omega_{min}^2}u=0.
\ea
In this case the kernel is modified to become
\ba
K(t)=-\int _{-\infty }^{+\infty }
\frac{d\omega }{(2\pi )}
\frac{{\omega _c}^4~\exp [-i\omega t]}{{\omega _c}^2
-i\gamma \sqrt{\omega ^2-{\omega_{min}^2}}}.
\ea
When $\omega _{min}$ is not too large, $\omega _{min} < \lambda =
\omega_c^2/\gamma$, there is an isolated pole
at $\omega =-i\Lambda $ where
\ba
\Lambda & = & \sqrt{\frac{{\omega _c}^4}{\gamma ^2}-{\omega_{min}^2}} =  \sqrt{\lambda^2-\omega _{min}^2}
\ea
and a branch cut on the real axis extending from
$\omega =-\omega _{min}$ to $\omega =+\omega _{min}.$
To obtain a causal kernel, we must take a contour passing
above the branch cut on the real axis. For $t>0,$ where the kernel
has support, we evaluate the contour by deforming it so that there are
two contributions
$K(t)=K_1(t)+K_2(t)$ (see Fig. \ref{fig:pole}(a)).
The first term originates from the contour encircling the
pole in the clockwise sense
\ba
K_1(t)=
-(\omega _c^2\lambda)\theta(t) \exp [-\Lambda t]
\frac {\sqrt{\Lambda ^2+{\omega_{min}^2} }}{\Lambda }.
\ea
The second originates from the contour encircling the branch cut
in the clockwise sense as indicated in Fig.\ref{fig:pole}(a).
\ba
K_2(t)=
-\omega_c^2\lambda \int _{-\omega _{min}}^{+\omega _{min}}
\frac{d\omega }{(2\pi )}
\exp [-i\omega t]\left[\frac{1}
{\sqrt{{\omega_{min}^2}-\omega ^2}+\lambda }-\frac{1}
{-\sqrt{{\omega_{min}^2}-\omega ^2}+\lambda }\right].
\ea

As $\omega_{min}$ approaches $\lambda = \omega_c^2/\gamma$, the pole on the
lower half-plane approaches $\omega = 0$. There is a second pole at
the mirror image $\omega = +i\Lambda$. However, this pole lies on the other
Riemann sheet hidden beyond the branch cut. When $\omega_{min}$ exceeds $\lambda =
\omega_c^2/\gamma$, these two poles collide and scatter at right angles so
that they lie on the real line on the branch cut (or just below after
the deformation in Fig. \ref{fig:pole}(b)). We can make both these poles disappear by deforming
the branch cut downward as indicated in fig. \ref{fig:pole}(c). The branch cut
has been pushed all the way to negative infinity.
Now the kernel integral may be expressed as
\ba
K(t) & = & -\omega^2_c\lambda\int_{C_3}\frac{d\omega }{(2\pi )}
\frac{\exp [-i\omega t]}{\lambda
-i\sqrt{\omega ^2-{\omega_{min}^2}}}-\omega^2_c\lambda\int_{C_4}\frac{d\omega }{(2\pi )}
\frac{\exp [-i\omega t]}{\lambda
-i \sqrt{\omega ^2-{\omega_{min}^2}}}
\ea
where the contours $C_3$ and $C_4$ are indicated in \ref{fig:pole}(c). We may
rewrite this as

\ba\label{qqq:kyes}
K(t) & = & -i \omega^2_c\lambda\int_0^\infty \frac{dy}{2\pi} \left( \frac{e^{-yt}e^{-i\omega_{min} t}}{\lambda-i\sqrt{y(-y-i2\omega_{min})}}-\frac{e^{-yt}e^{-i\omega_{min} t}}{\lambda+i\sqrt{y(-y-i2\omega_{min})}}\right)\cr
     & & - i \omega^2_c\lambda\int_0^\infty \frac{dy}{2\pi} \left( \frac{e^{-yt}e^{+i\omega_{min} t}}{\lambda-i\sqrt{y(-y+i2\omega_{min})}}-\frac{e^{-yt}e^{+i\omega_{min} t}}{\lambda+i\sqrt{y(-y+i2\omega_{min})}}\right).
   \ea
In fact this expression also holds for $\omega_{min} < \lambda =
\omega_c^2/\gamma$.
At large times the kernel
 (\ref{qqq:kyes}) behaves asymptotically as
\ba
K(t) &\stackrel{t\rightarrow\infty}{\approx}&
\mathcal{O}(1)\gamma\sqrt{\omega_{min}}{\cos\left(\omega_{min}t\right)-\sin\left(\omega_{min}t\right)\over
  t^{3/2}}.
\ea

We also compute the kernel in Fourier space. With the forcing term $f(t) =
f\cdot \exp [-i\omega t],$ the oscillator
equation is

\ba
-\omega ^2q(\omega )+
({\omega _0}^2+{\omega _c}^2)q(\omega )-
\frac{{\omega _c}^2\lambda}{\lambda-i\sqrt{\omega^2-\omega^2_{min}} }q(\omega )
& = & f(\omega ),\cr
\hat{G}^{-1}(\omega)q(\omega) & = & f(\omega).
\ea
The time average of the power transmitted to the oscillator (which is a
measure of the dissipation) is
\ba
\bar{P} & = & {1\over T}\int_0^T f(t)\dot{q}(t)dt,\cr
        & = & \omega\left(\Re [f] \Im [q] - \Im [f] \Re [q]\right),\cr
        & = & \vert f\vert^2 \omega \Im \left[\hat{G}(\omega)\right].
\ea
If the kernel has no imaginary part, the system exchanges energy but does not dissipate since
the average power vanishes. For $\omega < \omega_{min}$ the imaginary part of
the kernel vanishes. From $\hat{G}(\omega)$ we deduce the average dissipated power
\ba
\bar{P} & = & \vert f\vert^2  {\omega^2_c\lambda\omega
  \sqrt{\omega^2-\omega^2_{min}}\over
  \lambda^2\left(\omega_0^2-\omega^2\right)^2+\left(\omega^2-\omega^2_{min}\right)\left(\omega_0^2+\omega^2_c-\omega^2\right)^2}~~\qquad (\omega^2 > \omega_{min}^2).
\ea
The dissipated power is zero when
$\omega^2 \leq \omega_{min}^2$. One can understand qualitatively how the
behavior changes as $\omega _{min}$ is varied. The string may be regarded as a
high pass filter with threshold $\omega _{min}$. Only radiation of higher
frequency can escape down the string to infinity, thus leading to a net
flux of power away from the oscillator. At lower frequencies the
string is excited around the oscillator but in a localized way with
the amplitude decaying exponentially with distance from the oscillator since
the extra momentum $k = \sqrt{\omega^2-\omega^2_{min}}$ from the bulk wave equation
becomes imaginary.
 In this
case there is no dissipation in the long term because the string
vibrates in phase with the oscillator. Of course the coupling shifts the frequency
of the oscillator because the string adds both inertia and restoring
force to the oscillator.

For the brane-bulk system, the threshold is given by
\ba
 \omega^2_{min} = \lambda^2+\bar{\omega}^2
 \ea
where $\lambda$, $\bar{\omega}$ have been computed in
(\ref{eqnarray:omtilde}). For a weak coupling to the bulk ($\dot{H}\ll H^2$),
the frequency of the brane degree of freedom is only slightly displaced from its bare frequency
$\omega_0$ given in eqn. (\ref{eqnarray:param}). At subhorizon scales ($p\gg aH$),
\ba
0 < \omega_0^2 \approx {c_s^2 p^2\over a^2}+\mathcal{O}\left(\dot{H}\right) & < &
\omega^2_{min}\approx {p^2\over a^2}+\mathcal{O}\left(\dot{H}\right)
\ea
because $c_s^2\leq 1$. Therefore the dissipated power is exactly zero. In the
adiabatic approximation, valid on subhorizon scales, the fluid cannot
dissipate into the bulk because the natural frequency of the bulk mode at the
same wavenumber is too high to have a resonance.

\section{Discussion and conclusions}\label{sec:conclusion}

In this paper we presented analytic approaches to estimate the order of
magnitude of the dissipative effects in braneworld cosmology affecting the evolution of both scalar and tensor
perturbations on the brane. Because of the non-uniform expansion of the brane, the degrees of freedom on
the brane interact with the bulk gravitons, possibly leading to dissipation
from the four dimensional perspective. These effects are encoded in the $AdS$
bulk retarded propagator. From the dressed brane propagator, obtained by resumming the bulk backreaction effects at all order in the brane-bulk coupling, we may extract the effective dissipation rate of certain brane degrees of freedom. We assumed an adiabatic expansion on the brane ($\dot{H}\ll H^2$) in order to apply our
methods. 

Our analytic results provide more intuition on the previous
numerical simulations regarding braneworld slow-roll inflation \cite{scalinf3}. For the scalar perturbations with a slow-roll inflaton field on
the brane we obtained, without any other approximation, the local dissipation
rate of the inflaton perturbation due to its interaction with the bulk
gravitons, $\Gamma\sim
H\left(\dot{H}/H^2\right)\left(H\ell/\sqrt{1+H^2\ell^2}\right)$, which is linear in the slow-roll factor.  This is our main result and agrees with the numerical results obtained by Koyama \&
al. in \cite{scalinf3}. In the high-energy limit of braneworld inflation
($H\ell\gg 1$) the correction to standard inflation due to the coupling to
bulk metric perturbations is found to be linear in the slow-roll
factor as well. Moreover the dependence on $H\ell$ of the dissipation
rate correctly fits the behaviour of the slow-roll correction plotted in \cite{scalinf3}.

For the tensor perturbations we encounter purely nonlocal brane-bulk
interaction, that is why we simplified the backgound geometry: we explicitely neglected the
nonlocal processes on the brane due to backscattering of bulk gravitons in
curved $AdS$ assuming that  the dissipation is the dominant effect on the
brane by considering the near-brane limit of the bulk geometry (plateau potential). This approximation is well legitimized in the high-energy regime of braneworld inflation ($H\ell\gg 1$) because the bulk
curvature becomes negligeable compared to the acceleration of the
brane. We were able in this way to compute the slow-roll correction to tensor perturbations due to
the coupling of the zero mode with the continuum of bulk gravitons. We found a
small attenuation of the growing mode of the graviton bound state at superhorizon scales and at high
energy due to its interaction with the bulk modes and we obtained that this
correction is quadratic in the slow-roll factor, $\Gamma\sim
H(\dot{H}/H^2)^2$. This is a new result. The
inflaton field perturbation thus decays at a larger rate $\Gamma\sim H(\dot{H}/
H^2)$, linear in the adiabatic slow-roll parameter,
than the decay rate the graviton bound state $\Gamma\sim
H(\dot{H}/H^2)^2$, quadratic in the slow-roll parameter at high-energy.
This difference is the consequence of the locality of the interaction between
the inflaton and the bulk modes as opposed to the nonlocal interaction
between the graviton bound mode and the bulk modes.
We also found that an adiabatic perfect fluid on the brane does not dissipate into
the extra dimension at subhorizon scales because the minimum frequency of
the bulk modes at the same wavenumber as the fluid is too high to have a resonance. We encounter some
difficulties in applying analytic methods for the perfect fluid at superhorizon
scales because at such scales the expansion of the brane is no longer
adiabatic, at least when the fluid dominates the tension of the brane.

It would be interesting to go beyond the plateau potential approximation
by taking into account the inhomogeneity of the $AdS$ bulk and the diffraction of the
gravitons by the bulk which lead to nonlocal effects on the brane. The
nonlocality is encoded in the $AdS$ bulk propagator and depends strongly on
the bulk curvature. In that sense the exact expression of the $AdS$
propagator is necessary to compute the nonlocal effects. The
curved geometry of the bulk also requires the existence of metastable gravitons or
quasi-bound states on the brane even if the brane is static as it was found by Seahra \cite{seahra}. We
could compare the flux of the tunneling bulk gravitons with the flux of the
quasi-bound states. This will be the object of a future publication.
\newline

\flushleft{
{\Large {\bf Acknowledgments}}

I would like to thank Martin Bucher, Carla Carvalho, Christos Charmousis and Renaud Parentani for useful discussions.}


\begin{thebibliography}{99}
\bibitem{rs}
L.~Randall and R.~Sundrum, ``An Alternative to Compactification,"
Phys. Rev. Lett. 83, 4690 (1999); L.~Randall and R.~Sundrum, ``A Large Mass
Hierarchy From a Small Extra Dimension,'' Phys. Rev. Lett. 83, 3370 (1999).
\bibitem{gt}
J.~Garriga and T.~Tanaka, ``Gravity in the Randall-Sundrum Brane World,''
Phys. Rev. Lett. 84, 2778 (2000).
\bibitem{exp}
D.~J.~Kapner, T.~S.~Cook, E.~G.~Adelberger, J.~H.~Gundlach, B.~R.~Heckel,
C.~D.~Hoyle and H.~E.~Swanson, ``Tests of the Gravitational Inverse-Square Law
below the Dark-Energy length scale,'' Phys. Rev. Lett. 98:021101 (2007).
\bibitem{bdel}
P.~Bin\'etruy, C.~Deffayet, U.~Ellwanger and D.~Langlois, ``Brane Cosmological
Evolution in a Bulk With Cosmological Constant,'' Phys. Lett. B477, 285
(2000); P.~Bin\'etruy, C.~Deffayet and D.~Langlois, ``Nonconventional
Cosmology From a Brane Universe,'' Nucl. Phys. B565, 269 (2000); P.~Bowcock,
C.~Charmousis and R.~Gregory, ``General Brane Cosmologies and their Global
Space-Time Structure,'' Class. Quant. Grav. 17, 4745 (2000); T.~Shiromizu, K.~Maeda and M.~Sasaki, ``The Einstein Equations on the 3-Brane
World,'' Phys. Rev. D 62, 024012 (2000).
\bibitem{lan}
J.~Garriga and M.~Sasaki, "Brane-World Creation and Black Holes," Phys. Rev. D62:043523 (2000);
D.~Langlois, R.~Maartens and D.~Wands , ``Gravitational Waves from Inflation on
the Brane,'' Phys. Lett. B489, 259 (2000); A.~V.~Frolov and L.~Kofman, ``Gravitational Waves from Brane World
Inflation,'' hep-th/0209133.
\bibitem{deff}
C.~Deffayet, ``On Brane World Cosmological Perturbations,'' Phys. Rev.
D66, 103504 (2002); C.~Deffayet, ``Note on the Well-Posedness of Scalar Brane
World Cosmological Perturbations,'' Phys. Rev.
D71, 023520 (2005).
\bibitem{pert}
D.~Langlois, ``Brane Cosmological Perturbations,'' Phys. Rev. D62, 126012 (2000);  R.~Easther, D.~Langlois, R.~Maartens and D.~Wands, ``Evolution of Gravitational Waves in Randall-Sundrum Cosmology,'' JCAP 0310, 014
(2003); A.~Riazuelo,
F.~Vernizzi, D.~Steer and R.~Durrer, ``Gauge Invariant Cosmological
Perturbation Theory for Braneworlds,'' hep-th/0205220; D.~Langlois, L.~Sorbo, M.~Rodriguez-Martinez, ``Cosmology of
a Brane Radiating Gravitons Into the Extra Dimension,'' Phys. Rev. Lett. 89, 171301 (2002); D.~Langlois, R.~Maartens, M.~Sasaki, D.~Wands, ``Large Scale Cosmological
Perturbations on the Brane,'' Phys. Rev. D63, 084009 (2001); R.~A.~Battye, C.~van de Bruck and A.~Mennim, ``Cosmological Tensor
Perturbations in the Randall-Sundrum Model: Evolution in the Near-Brane
Limit,'' Phys. Rev. D69, 064040 (2004).
\bibitem{pert2}
S.~Mukohyama, ``Gauge Invariant Gravitational Perturbations of Maximally
Symmetric Spacetimes,'' Phys. Rev. D62, 084015 (2000); S.~Mukohyama,
``Perturbation of Junction Condition and Doubly Gauge Invariant Variables,''
Class. Quant. Grav. 17, 4777 (2000); S.~Mukohyama, ``Integro-differential
Equation for Brane World Cosmological Perturbations,'' Phys. Rev. D64:064006
(2001) (Erratum-ibid. D66, 049902 (2002)); S.~Mukohyama, ``Doubly Covariant
Action Principle of Singular Hypersurfaces in General Relativity and
Scalar-Tensor Theories,'' Phys. Rev. D65, 024028 (2002); S.~Mukohyama,
``Doubly Covariant Gauge Invariant Formalism of Braneworld Cosmological
Perturbations,'' hep-th/0202100; S.~Mukohyama, ``Nonlocality
as an Essential Feature of Brane Worlds,'' Prog. Theor. Phys. Suppl. 148, 121
(2003).
\bibitem{jap}
H.~Kodama, A.~Ishibashi and O.~Seto, ``Brane World Cosmology: Gauge-Invariant
Formalism for Perturbation,'' Phys. Rev. D62, 064022 (2000).
\bibitem{bridg}
H.~A.~Bridgman, K.~A.~Malik and D.~Wands, ``Cosmological Perturbations in the
Bulk and on the Brane,'' Phys. Rev. D65, 043502 (2002).
\bibitem{pert3}
D.~S.~Gorbunov, V.~A.~Rubakov and S.~M.~Sibiryakov, ``Gravity Waves From
Inflating Brane or Mirrors Moving in AdS5,'' JHEP 0110, 015 (2001);
T.~Kobayashi, H.~Kudoh and T.~Tanaka, `` Primordial Gravitational Waves in
Inflationary Brane World,'' Phys. Rev. D68:044025 (2003).
\bibitem{der}
N.~Deruelle, ``Cosmological Perturbations of an Expanding Brane in an Anti-de
Sitter Bulk: A Short Review,'' Astrophys. Space Sci. 283:619-626 (2003).
\bibitem{scalbc}
K.~Koyama, ``Late Time Behaviour of Cosmological Perturbations in a
Single-Brane Model,'' JCAP0409,010 (2004); K.~Koyama, D.~Langlois, R.~Maartens
and D.~Wands, ``Scalar Perturbations from
Brane-World Inflation,'' JCAP0411,002 (2004).
\bibitem{num0}
T.~Hiramatsu, K.~Koyama and A.~Taruya, ``Evolution of gravitational waves
from inflationary brane world : numerical study of high-energy effects,''
Phys.Lett.B578:269-275 (2004).
\bibitem{num1}
T.~Hiramatsu, K.~Koyama and A.~Taruya, ``Evolution of gravitational waves in
the high-energy regime of brane-world cosmology,'' Phys.Lett.B609:133-142
(2005).
\bibitem{scalinf}
T.~Hiramatsu and K.~Koyama , `` Numerical Study of Curvature Perturbations in a
Brane-World Inflation at High-Energies,'' JCAP 0612:009 (2006).
\bibitem{num2}
T.~Hiramatsu, ``High-energy effects on the spectrum of inflationary
gravitational wave background in braneworld cosmology,'' Phys.Rev.D73:084008 (2006).
\bibitem{num3}
T.~Kobayashi and T.~Tanaka, ``Quantum-mechanical generation of gravitational
waves in braneworld,'' Phys.Rev.D71:124028 (2005).
\bibitem{num4}
T.~Kobayashi, ``Initial Kaluza-Klein Fluctuations and Inflationary
Gravitational Waves in Braneworld Cosmology,'' Phys. Rev. D73:124031
(2006).
\bibitem{num5}
S.~S.~Seahra, ``Gravitational Waves and Cosmological Braneworlds: A
Characteristic Evolution Scheme,'' Phys. Rev. D74:044010
(2006).

\bibitem{scal}
A.~Cardoso, T.~Hiramatsu, K.~Koyama and S.~S.~Seahra, ``Scalar Perturbations
in Braneworld Cosmology,'' arXiv:0705.1685[astro-ph].

\bibitem{scalinf2}
 K.~Koyama, A.~Mennim, V.~A.~Rubakov,
D.~Wands and T.~Hiramatsu, ``Primordial Perturbations from Slow-roll Inflation
on a Brane,'' JCAP 0704:001 (2007).
\bibitem{scalinf3}
K.~Koyama, A.~Mennim and D.~Wands, ``Brane-world inflation: Slow-roll
corrections to the spectral index,'' Phys.Rev.D77:021501 (2008).
\bibitem{trodden}
G.~D.~Starkman, D.~Stojkovic and M.~Trodden, ``Homogeneity, Flatness and
'Large' Extra Dimensions,''  Phys.Rev.Lett.87:231303 (2001); G.~D.~Starkman, D.~Stojkovic and M.~Trodden, ``Large Extra Dimensions and
Cosmological Problems,'' Phys.Rev.D63:103511 (2001).
\bibitem{bucher}
 M.~Bucher, ``A Brane World Universe from Colliding Bubbles,''
 Phys.Lett.B530:1-9 (2002).
\bibitem{bbc}
P.~Binetruy, M.~Bucher and C.~Carvalho, ``Models for the Brane Bulk
Interaction: Toward Understanding Brane World Cosmological Perturbations,''
Phys. Rev. D70:043509 (2004); M.~Bucher and C.~Carvalho, `` Linearized Israel
Matching Conditions for Cosmological Perturbations in a Moving Brane
Background,'' Phys. Rev. D71:083511 (2005).
\bibitem{reso}
K.~Koyama, A.~Mennim and D.~Wands, ``Coupled boundary and bulk fields in
Anti-de Sitter spacetime,'' Phys.Rev.D72:064001 (2005);
A.~Cardoso, K.~Koyama, A.~Mennim,
S.~S.~Seahra and D.~Wands, ``Coupled Bulk and Brane Fields about a de Sitter
Brane,'' Phys. Rev. D75:084002 (2007).
\bibitem{seahra}
S.~S.~Seahra, ``Ringing the Randall-Sundrum braneworld:Metastable gravity wave
bound states,'' Phys.Rev.D72:066002 (2005); 
S.~S.~Seahra, ``Metastable Massive Gravitons from an Infinite Extra
Dimension,'' Int. J. Mod. Phys. D14:2279-2284 (2005).
\end{thebibliography}
\end{document}